\pdfoutput=1
\documentclass[11pt]{article}

\usepackage{epsfig,latexsym,amsfonts,amsmath,amsthm,amssymb,amsbsy,multirow,slashed,wasysym,textcomp,subfigure,hyperref,wrapfig,datetime,comment,mathtools,cancel,cite,mathrsfs,footmisc}
\usepackage{orcidlink}
\usepackage{datetime2}
\usepackage[font={footnotesize,sl},bf]{caption}

\usepackage{setspace}
\usepackage{tikz}
\usetikzlibrary{calc}
\usetikzlibrary{decorations.pathreplacing}
\usetikzlibrary{decorations.pathmorphing}
\usetikzlibrary{decorations.markings}
\usetikzlibrary{arrows.meta}
\usetikzlibrary{positioning}

\usepackage{xcolor}
\definecolor{oldmauve}{rgb}{0.4, 0.19, 0.28}
\definecolor{pansypurple}{rgb}{0.47, 0.09, 0.29}
\definecolor{burgundy}{rgb}{0.5, 0.0, 0.13}
\definecolor{carminepink}{rgb}{0.92, 0.3, 0.26}
\definecolor{blue(pigment)}{rgb}{0.2, 0.2, 0.6}
\definecolor{darkseagreen}{rgb}{0.56, 0.74, 0.56}
\definecolor{darkspringgreen}{rgb}{0.09, 0.45, 0.27}
\definecolor{ceruleanblue}{rgb}{0.16, 0.32, 0.75}

\newcommand{\goo}{g_{tt}^{(2)}}

\newcommand{\li}{\left(}
\newcommand{\ri}{\right)}

\newcommand{\eq}[2]{\begin{equation} #1 \label{#2} \end{equation}}

\DeclareMathOperator{\extdm}{d}
\newcommand{\extd}{\extdm \!}

\numberwithin{equation}{section}

\def\bea{\begin{eqnarray}}
\def\eea{\end{eqnarray}}
\newcommand{\beq}{\begin{eqnarray}}
\newcommand{\eqq}{\end{eqnarray}}
 \newcommand{\badat}{\begin{alignedat}}
 \newcommand{\eadat}{\end{alignedat}}

\newcommand{\eal}[1]{\be \begin{aligned} #1 \end{aligned}\end{equation}} 
\newcommand{\eqn}[1]{\be #1 \end{equation}} 
\newcommand{\eqa}[1]{\bea  #1\end{eqnarray}}

\long\def\new#1\endnew{{\bf #1}}		
\long\def\del#1\enddel{}

\def\del{\partial}

\newcommand{\be}{\begin{eqnarray}}
\newcommand{\en}{\end{eqnarray}}

\newcommand{\mat}{{\cal M}} 

\numberwithin{equation}{section} 

\begin{document}


\begin{titlepage}
\DefineFNsymbols*{AdSbath}[math]{\blacktriangleleft\blacktriangleright\blacklozenge} 
\setfnsymbol{AdSbath}
\renewcommand{\thefootnote}{\fnsymbol{footnote}}
  \thispagestyle{empty}

 \begin{flushright}
 TUW--23--06
 \end{flushright}

  \begin{center}  
{\LARGE\textbf{One-loop partition function of gravity}}
\vskip0.2cm
{\LARGE\textbf{with leaky boundary conditions}}

\vskip1cm 
Daniel Grumiller,\footnote{\orcidlink{0000-0001-7980-5394}\fontsize{8pt}{10pt}\selectfont\ \href{mailto:grumil@hep.itp.tuwien.ac.at}{grumil@hep.itp.tuwien.ac.at}} 
Romain Ruzziconi,\footnote{\orcidlink{0000-0002-4359-7586}\fontsize{8pt}{10pt}\selectfont\ \href{mailto:Romain.Ruzziconi@maths.ox.ac.uk}{romain.ruzziconi@maths.ox.ac.uk}} and
Céline Zwikel~\footnote{\orcidlink{0000-0002-6582-6351}\fontsize{8pt}{10pt}\selectfont\ \href{mailto:czwikel@perimeterinstitute.ca}{czwikel@perimeterinstitute.ca}}
\vskip0.5cm

\normalsize
\medskip

$^\blacktriangleleft$\textit{Institute for Theoretical Physics, TU Wien\\ Wiedner Hauptstrasse~8-10/136 A-1040 Vienna, Austria}
\\ \smallskip
$^\blacktriangleleft$\textit{Theoretical Sciences Visiting Program, Okinawa Institute of Science and
Technology \\ Graduate University, Onna, 904-0495, Japan}
\\ \smallskip
$^\blacktriangleright$\textit{Mathematical Institute, University of Oxford, \\ Andrew Wiles Building, Radcliffe Observatory Quarter, \\
Woodstock Road, Oxford, OX2 6GG, UK} \\ \smallskip
$^\blacklozenge$\textit{Perimeter Institute for Theoretical Physics\\31 Caroline Street North, Waterloo, Ontario, Canada N2L 2Y5} 
\end{center}

\vskip0.5cm

\begin{abstract}
Leaky boundary conditions in asymptotically AdS spacetimes are relevant to discuss black hole evaporation and the evolution of the Page curve via the island formula. We explore the consequences of leaky boundary conditions on the one-loop partition function of gravity. We focus on JT gravity minimally coupled to a scalar field whose normalizable and non-normalizable modes are both turned on, allowing for leakiness through the AdS boundary. Classically, this yields a flux-balance law relating the scalar news to the time derivative of the mass. Semi-classically, we argue that the usual diffeomorphism-invariant measure is ill-defined, suggesting that the area-non-preserving diffeomorphisms are broken at one loop. We calculate the associated anomaly and its implication on the gravitational Gauss law. Finally, we generalize our arguments to higher dimensions and dS.
\end{abstract}

\renewcommand{\thefootnote}{\arabic{footnote}}
\setcounter{footnote}{0}
\end{titlepage}

\tableofcontents


\section{Introduction}

\begin{figure}[bth]
\begin{center}
\begin{tikzpicture}[scale=0.6]
	\tiny
\draw[red,opacity=0] (-3,-8.6) -- (10,-8.6) -- (10,1.5) -- (-3,1.5) -- cycle;
\def\dx{.2};
\def\dy{.4};
\def\shiftx{9.6};
\def\shifty{-7};
\coordinate (A) at (-2,-7+\dy);
\coordinate (B) at (2,-7+\dy);
\coordinate (C) at (2,-\dy);
\coordinate (D) at (-2,-\dy);

\coordinate (E) at (-2+\shiftx,\dy+\shifty);
\coordinate (F) at (2+\shiftx,\dy+\shifty);
\coordinate (G) at (2+\shiftx,7-\dy+\shifty);
\coordinate (H) at (-2+\shiftx,7-\dy+\shifty);

\draw[red,opacity=0] ($(A)+(-1,-1)$) -- ($(D)+(-1,1)$) -- ($(G)+(1,1)$) -- ($(F)+(1,-1)$) -- cycle;

\draw (B) arc[x radius=2, y radius=0.5, start angle=0, end angle=-180];
\draw [densely dashed] (B) arc[x radius=2, y radius=0.5, start angle=0, end angle=180];
\draw (A) -- (D);
\draw (B) -- (C);
\draw (C) arc[x radius=2, y radius=0.5, start angle=0, end angle=-360];

\draw (E) -- (H);
\draw (F) -- (G);
\fill [white] (F) arc[x radius=2, y radius=0.5, start angle=0, end angle=360];
\draw (F) arc[x radius=2, y radius=0.5, start angle=0, end angle=-180];
\draw [densely dashed] (F) arc[x radius=2, y radius=0.5, start angle=0, end angle=180];
\draw (G) arc[x radius=2, y radius=0.5, start angle=0, end angle=360];

\def\decal{0.1};

\path [-{Latex[width=1mm]},draw=red] (0.5,-5.5) -- (2,-4) -- (0,-2);
\path [-{Latex[width=1mm]},draw=red] (0.25,-5.25) -- (2,-3.5) -- (0.25,-1.75);
\path [-{Latex[width=1mm]},draw=red] (0,-5) -- (2,-3) -- (0.5,-1.5);
\path [blue] (-2,-4.5) edge[bend left=30] (0,-4.5) -- (0,-4.5) edge[bend right=30] (2,-4.5);
\draw (-1,-4) node[above,blue,inner sep=2pt]{$\Sigma$};

\path [-{Latex[width=1mm]},draw=red] (0.5+\shiftx,1.5+\shifty) -- (2+\shiftx,3+\shifty) -- (3.25+\shiftx,4.25+\shifty);
\path [-{Latex[width=1mm]},draw=red] (0.25+\shiftx,1.75+\shifty) -- (2+\shiftx,3.5+\shifty) -- (3+\shiftx,4.5+\shifty);
\path [-{Latex[width=1mm]},draw=red] (0+\shiftx,2+\shifty) -- (2+\shiftx,4+\shifty) -- (2.75+\shiftx,4.75+\shifty);
\path [blue] (-2+\shiftx,2.5+\shifty) edge[bend left=30] (0+\shiftx,2.5+\shifty) -- (0+\shiftx,2.5+\shifty) edge[bend right=30] (2+\shiftx,2.5+\shifty);
\draw (-1+\shiftx,3+\shifty) node[above,blue,inner sep=2pt]{$\Sigma$};

\coordinate (C1) at ($(F)!0.9!(G)$);
\coordinate (C2) at ($(F)!0.5!(G)$);
\coordinate (C3) at ($(F)!0.1!(G)$);
\coordinate (D1) at ($(C1)+(3*\decal,0)$);
\coordinate (D2) at ($(C2)+(2.5+3*\decal,0)$);
\coordinate (D3) at ($(C3)+(3*\decal,0)$);

\draw ($(A)!0.5!(B)-(0,0.5)$) node[fill=white,inner sep=2pt]{$i^-_{\text{AdS}}$};
\draw ($(G)!0.5!(H)+(0,0.5)$) node[fill=white,inner sep=2pt]{$i^+_{\text{AdS}}$};
\draw ($(E)!0.5!(F)-(0,0.5)$) node[fill=white,inner sep=2pt]{$i^-_{\text{AdS}}$};
\draw ($(C)!0.5!(D)+(0,0.5)$) node[fill=white,inner sep=2pt]{$i^+_{\text{AdS}}$};
\draw ($(B)!0.8!(C)$) node[right]{$\mathscr I_{\text{AdS}}$};
\end{tikzpicture}
\caption{Left: Penrose diagram of AdS with reflective boundary conditions (closed system). Red lines represent propagating massless modes. Right: Penrose diagram of AdS with leaky boundary conditions (open system). External bath can be glued at AdS boundary to collect radiation.}
\label{fig:leaks}
\end{center}
\end{figure}
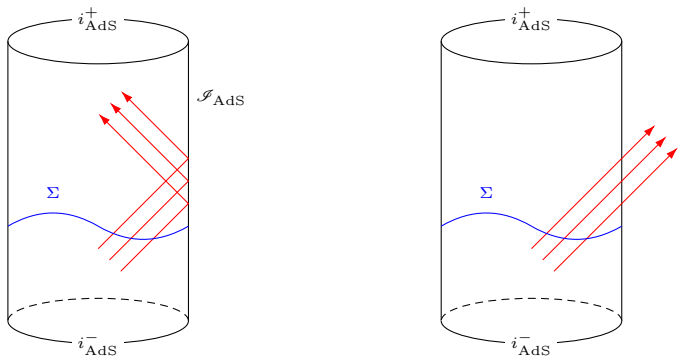

Large black holes in asymptotically AdS spacetimes are thermodynamically stable if one imposes reflective boundary conditions, such as Dirichlet or Neumann. This is so because they put the black hole in a Hartle--Hawking state since the Hawking radiation bounces off the asymptotic AdS boundary and gets reabsorbed by the black hole, which thus never evaporates. The natural way to evade this boring situation and allow for complete black hole evaporation is to couple AdS spacetime with an external bath that acts as a reservoir for the Hawking radiation (see Figure~\ref{fig:leaks}). This construction is the starting set-up for the recent discussions on the Page curve using the island formula in AdS \cite{Almheiri:2019yqk , Almheiri:2019qdq, Almheiri:2020cfm}. Remarkably, it was shown in \cite{Geng:2020qvw, Geng:2021hlu} that this coupling to the external bath has drastic implications at the quantum level, such as the breaking of the gravitational constraint via a spontaneous symmetry breaking, interpreted as gravitons becoming massive via a St\"uckelberg mechanism \cite{Porrati:2001db}. 


In practice, the coupling with the thermal bath can be described by considering leaky boundary conditions, allowing for the exchange of energy and matter through the permeable boundary. These boundary conditions were studied in detail in \cite{Compere:2019bua,Compere:2020lrt,Fiorucci:2020xto} for classical pure general relativity in arbitrary dimensions. The upshot is that the leaks of gravitational waves through the boundary yield non-conservation of the asymptotic charges, which can be described by flux-balance laws (instead of the standard conservation laws associated with reflective boundary conditions). This approach is non-standard in AdS/CFT where one instead is interested in describing gravity in a closed system (however, for aspects of non-equilibrium AdS/CFT, see e.g. \cite{Hubeny:2010ry,Bakas:2014kfa,Skenderis:2017dnh,Ciambelli:2017wou,Liu:2018crr,Kundu:2019ull}).  Leaky boundary conditions can be considered in other types of asymptotics, such as asymptotically flat spacetimes \cite{Wald:1999wa,Barnich:2011mi,Compere:2018ylh,Compere:2019gft,Freidel:2021fxf,Donnay:2022wvx} where they yield the famous Bondi mass loss formula \cite{Trautman:1958zdi,Bondi:1962}, at the future boundary of asymptotically dS spacetime \cite{Anninos:2010zf,Compere:2019bua,Compere:2020lrt,Poole:2021avh,Bonga:2023eml,Compere:2023ktn,Geiller:2022vto}, as well as at boundaries located at finite distance \cite{Chandrasekaran:2018aop,Adami:2020amw,Adami:2020ugu,Adami:2021sko,Adami:2021nnf,Adami:2022ktn,Freidel:2022bai,Chandrasekaran:2023vzb,Odak:2023pga}.

In two and three spacetime dimensions, there are no propagating degrees of freedom in pure gravity, which makes the interpretation of leaky boundary conditions less straightforward (see, however, \cite{Alessio:2020ioh,Ruzziconi:2020wrb,Geiller:2021vpg,Campoleoni:2022wmf,McNees:2023tus,Ciambelli:2023ott} where they are interpreted in terms of the boundary geometry). Nevertheless, one can implement leaky boundary conditions by simply coupling gravity with a matter field (see, e.g., \cite{Barnich:2015jua,Bosma:2023sxn} for an example in three-dimensional Einstein--Maxwell theory). In this work, we focus on JT gravity in asymptotically AdS spacetimes minimally coupled to a massless scalar field whose associated propagating degrees of freedom can leak through the conformal boundary. This constitutes the typical setup to discuss the black hole evaporation and the evolution of the Page curve via the island formula. 

While this system is perfectly under control at the classical level, we explain why the quantization is a subtle issue, which requires enlarging the Hilbert space to include modes transmitted through the asymptotic boundary. Notably, we observe that the standard diffeomorphism-invariant measure for the scalar field is ill-defined due to the weak falloff of the latter at infinity implied by the inclusion of the extra modes. We explore various alternative quantization schemes to resolve this issue, and are led to the following observation: There is no natural, consistent, local, diffeomorphism-invariant path integral measure for the scalar field. In other words, leaky boundary conditions break parts of the diffeomorphisms in the one-loop partition function of gravity. 

We investigate in detail an alternative quantization that involves a Weyl-invariant measure but breaks diffeomorphisms that are not area-preserving \cite{Karakhanian:1994gs,Jackiw:1995qh}. Hence, the main feature of this quantization scheme is to transfer the usual Weyl anomaly to the non-area-preserving diffeomorphisms. We derive the diffeomorphism anomaly arising at one loop and show that it violates the gravitational constraint. This phenomenon is reminiscent of the higher-dimensional case \cite{Geng:2020qvw, Geng:2021hlu} where the gravitational constraint (or gravitational ``Gauss law'') is broken at the quantum level. Since we are in two dimensions (2d), there is no interpretation of this breaking as bulk gravitons becoming massive. However, this interpretation might be recovered by adding additional degrees of freedom into the theory and re-interpreting the diffeomorphism anomaly as a spontaneous symmetry breaking via the St\"uckelberg mechanism. 

Finally, let us emphasize that the present work takes the point of view of AdS as an open system and does not assume anything about the specific nature of the bath. This constitutes a toy model to address the challenging question of understanding subsystems in gravity \cite{Donnelly:2016auv,Speranza:2017gxd,Witten:2018zxz,Kirklin:2018gcl,Camps:2018wjf,Camps:2019opl,Bahiru:2022oas,Chandrasekaran:2022cip,Bahiru:2023zlc,Jensen:2023yxy,Folkestad:2023cze,Witten:2023xze}. We expect that the closure of the system by adding an explicit bath might shed some light on the exotic features we are discussing here, and we refer to \cite{Geng:2023ynk} for a recent work along these lines. 

This work is organized as follows. In section \ref{sec:Leaky boundary conditions}, we introduce leaky boundary conditions and derive the classical flux-balance law. In section \ref{sec:lbcdiffeoanomaly}, we study the leaky model semi-classically and point out the issues that lead to the diffeomorphism anomaly. In section \ref{sec:Local and Weyl invariant measure}, we propose employing a local Weyl-invariant path integral measure that avoids the difficulties of the previous section. In section \ref{sec:5}, we discuss the consequences of leakiness for the Gauss law. In section \ref{sec:7}, we put our results into the context of recent literature and generalize the discussion to higher dimensions and dS. In appendix \ref{sec:Derivation of the asymptotic charges}, we provide a detailed derivation of the asymptotic charges. In appendix \ref{sec:holography}, we give a holographic interpretation of some of our results. In appendix \ref{sec:6}, we address an alternative dilaton-dependent measure. In appendix \ref{app:IyerWaldeffectiveaction}, we provide more details on the Gauss law breaking.

\section{Leaky boundary conditions and flux-balance laws}
\label{sec:Leaky boundary conditions}

In this section, we present the main set-up of this paper, namely JT gravity \cite{Teitelboim:1983ux,Jackiw:1984} (in the linear dilaton sector) minimally coupled to a massless scalar field in AdS$_2$ with leaky boundary conditions. See \cite{Kummer:1996hy,Almheiri:2014cka,Engelsoy:2016xyb,Pedraza:2021cvx,Ecker:2022vkr,Mertens:2022irh,Bak:2023zkk} for earlier work on (semi-classical) scalar fields coupled to JT gravity. We present a detailed analysis of the asymptotic symmetries in the presence of leaks and explain how the flux-balance laws, which control the non-conservation of the charges, can be deduced from the symmetries. 

\subsection{Solution space}
The action 
\begin{equation}
\begin{split}
S[g_{\mu\nu},X, \psi]=\frac{1}{16\pi G}\int_{\mathscr M}\!\!\extd^2x \sqrt{-g}\, X\,\big(R - 2 \Lambda\big)+ \int_{\partial \mathscr M}\!\!\!\!\extd x \, L_b + \frac{1}{2} \int_{\mathscr M}\!\!\extd^2 x \sqrt{-g}\,g^{\mu\nu}(\partial_\mu\psi)(\partial_\nu\psi)
\end{split}\label{renormalized action}
\end{equation}  
functionally depends on the (Lorentzian) metric $g_{\mu\nu}$, the dilaton $X$, and the matter scalar field $\psi$. The manifold $\mathscr M$ is a disk when we continue to Euclidean signature, with boundary $\partial \mathscr M$. The gravitational coupling is positive, $G>0$, and the other parameter in the action is negative, $\Lambda=-\frac{1}{\ell^2}$. The first term on the right-hand side is the bulk JT gravity action, the second is a boundary term ensuring that the variational principle is well-defined for the gravitational sector (see \eqref{renormalized action 2} or (4.1) of \cite{Grumiller:2017qao} for the explicit expression of $L_b$), and the third term is a free massless scalar field minimally coupled to the gravitational field. The equations of motion for the action \eqref{renormalized action}, 
\begin{equation}
  \nabla_\mu \nabla_\nu X - g_{\mu\nu} \nabla^2 X + \frac{1}{\ell^2}\, g_{\mu\nu} X  = 8\pi G\, T_{\mu\nu}  \qquad\qquad
R = -\frac{2}{\ell^2} \qquad\qquad \nabla^2\psi = 0
\label{EOM matter}
\end{equation}
guarantee that all solutions are locally AdS$_2$. The stress-energy tensor therein,
\eq{
T_{\mu\nu} = (\partial_\mu \psi) (\partial_\nu \psi) - \frac{1}{2} g_{\mu\nu} g^{\alpha\beta} (\partial_\alpha \psi) (\partial_\beta \psi)
}{eq:angelinajolie}
classically is conserved
\eq{
\nabla^\mu T_{\mu\nu} = 0
}{eq:conserved}
and traceless
\eq{
{T^\mu}_\mu = 0
}{eq:traceless}
which is associated, respectively, with classical diffeomorphism- and Weyl-invariance of the scalar field action in \eqref{renormalized action}.

We impose the Fefferman--Graham gauge 
\begin{equation}\label{FGgauge}
    g_{\rho \rho}=\frac{\ell^2}{\rho^2}\qquad \qquad g_{t\rho}=0 
\end{equation} 
where $\rho$ is the radial coordinate ($\rho = 0$ at the boundary and $\rho > 0$ in the bulk) and $t$ the time. We use intrinsic 2d units, where  $\ell, t$ and $1/\rho$ have length dimension, while $X$ and $G$ are dimensionless.\footnote{%
From these choices, we deduce the length dimensions of all other quantities: $T_{\rho\rho}$: length squared; $N,X^-$: length; $T_{\rho t},J$: dimensionless; $X^+$: inverse length; $T_{tt}, C, M$: inverse length squared.
}
Fixing Dirichlet boundary conditions for the metric, 
\begin{equation}
\label{dirichlet BC}
    g_{tt} = - \frac{1}{\rho^2\ell^2} + \mathcal{O}(1)
\end{equation} 
solving the equations of motion  \eqref{EOM matter},  we find 
\begin{subequations}
\label{falloff conditions}
  \begin{align}
X &= \frac1{\rho}\,X^+(t)  + \rho\,\frac{\ell^2}{2X^+(t)}\li  M(t)-\frac{\ell^2}2\li X^+(t)'\ri^2\ri +\mathcal O(\rho^2)\\ \label{falloff conditionsmetric}
g_{tt} &= -\frac1{\ell^2} \left[ \frac{1}{\rho} -\rho\,\frac{\ell^2}{2X^+(t)^2} \left( M(t)-\frac{\ell^2}{2}\li X^+(t)'\ri^2+\ell^2 X^+\,X^+(t)'' \right) \right]^2 \\
    \psi &= J(t) + \rho\,N(t) +\mathcal O(\rho^2) 
\end{align}
\end{subequations}
where prime denotes derivatives with respect to time $t$, and we assume $X^+(t)\neq 0$ for all times. In appendix \ref{sec:holography}, we give a holographic interpretation of the expansion coefficients above. For completeness, we display the stress-energy tensor falloff,
\eq{
    T_{\rho\rho} = \frac12 \left(N^2+\ell^4\,(J')^2 \right) +\mathcal O(\rho) \,,\quad 
    T_{\rho t}  = N \, J'+\mathcal O(\rho) \,,\quad T_{tt}=\frac1{2\ell^4}\left(N^2+\ell^4 (J')^2 \right)+\mathcal O(\rho) \,.
}{eq:bulkT}
As in \cite{Grumiller:2017qao,Ruzziconi:2020wrb}, we allow for fluctuations $X^+(t)\neq 0$ of the leading order in the dilaton. The quantities $J(t)$ and $N(t)$ correspond to the non-normalizable and normalizable modes of the scalar field $\psi$, respectively. In addition, the parameter $M$, which coincides with the leading order of the Casimir function
\eq{
 C\equiv -\frac{(\nabla X)^2}2 + \frac{X^2}{2\ell^2} =\frac2{\ell^2}\left( X^+\,X^-+ \frac {\ell^4}4 {\left((X^+)'\right)^2}\right)+ \mathcal O(\rho) = M + \mathcal O(\rho)
}{eq:C} 
corresponds to the mass, as can be checked on the particular case of JT black hole solutions. In the absence of matter, the latter are given by
\eq{
\extd s^2 = \frac{\ell^2}{\rho^2}\,\extd\rho^2 - \left(\frac{1}{\rho\ell}-\rho\ell\,\frac{\ell^2M}{2}\right)^2\,\extd t^2\qquad\qquad X = \frac{1}{\rho\ell} + \rho\ell\,\frac{\ell^2M}{2}\,.
}{eq:JTsol}

The equations of motion \eqref{EOM matter} imply the flux-balance law
\begin{equation}
\boxed{
\phantom{\Big(}
    \frac{\extd M}{\extd t}  = \frac{8\pi G\,X^+}{\ell^2}\,N\,J'
\phantom{\Big)}    
    }
    \label{mass change}
\end{equation}  
relating the time derivative of the mass to a bilinear in the scalar field. We need to turn on a time-dependent non-normalizable mode $J(t)$ to obtain a non-vanishing flux on the right-hand side. Here, $J(t)$ can be interpreted as a boundary source describing an exchange of energy between AdS$_2$ and an external bath (see Figure \ref{fig:leaks}). By analogy with the case of asymptotically flat spacetime \cite{Trautman:1958zdi,Bondi:1962,Wald:1999wa,Barnich:2011mi}, we call $J'(t)$ the scalar news. Once $J(t)$ is turned off or becomes constant, there is no leak through the conformal boundary, and we are back to the standard situation where the mass is conserved in time.  

To summarize, the solution space is parametrized by three functions of the time coordinate, $X^+, N, J$, and by the function $M$ whose evolution is constrained by the flux-balance law \eqref{mass change}.

The non-normalizable modes of the scalar field correspond to those transmitted through the conformal boundary of AdS$_2$, while the normalizable modes are those that are reflected. Indeed, the most general solution to the 2d massless Klein-Gordon equation in \eqref{EOM matter} is given by the sum of holomorphic and anti-holomorphic parts, $\psi = f_+(t+\rho)+f_-(t-\rho) = f_+(t) + f_-(t) + \rho\,\big(f_+^\prime(t)-f_-^\prime(t)\big) + \dots$. The leading order term corresponds to the non-normalizable mode and the first subleading term to the normalizable mode; further subleading terms are not displayed and do not play a major role. We can also decompose into Fourier modes, $f_\pm = \sum_n f_n^\pm e^{in(t\pm\rho)}$. The non-normalizable modes then obey the condition $f_n^+=f_n^-$ and the normalizable modes obey the condition $f_n^+=-f_n^-$ which are recognized, respectively, as transmissive and reflective boundary conditions on the Fourier modes.

\subsection{Asymptotic symmetries}\label{sec:ASA}

Spacetime diffeomorphisms preserving the Fefferman-Graham gauge \eqref{FGgauge} are generated by
\eq{
    \xi^\rho = \sigma(t)\,\rho\qquad \qquad \xi^t 
    = f(t)+\frac{\ell^4\rho ^2\,\sigma(t)^\prime}{2-\ell^2 \rho ^2 \goo(t)} = f(t)+ \frac{\ell^4\rho^2}{2}\,\sigma(t) ^\prime+\mathcal O\li\rho^4\ri\,.
}{residual gauge diffeos}
Here, $\sigma(t)$ is the parameter for the Penrose--Brown--Henneaux diffeomorphisms \cite{penrosespinorsII,Brown:1986nw,Imbimbo:1999bj} inducing boundary Weyl rescalings, and $f(t)$ is the parameter for the tangential diffeomorphisms leading to re\-pa\-ra\-metri\-za\-tion symmetries at the boundary. Preserving Dirichlet boundary conditions for the metric \eqref{dirichlet BC} implies 
\eq{
   \sigma=f^\prime\,.
}{constraint parameters}
One can easily check that the above diffeomorphisms preserve the falloffs of the dilaton, the metric, and the matter field in \eqref{falloff conditions}. Assuming no field dependence in the symmetry parameters, the asymptotic symmetry algebra 
\eq{
    \big[\xi(f_1),\,\xi(f_2)\big]=\xi\big(f_1\,f^\prime_2-f_2\,f_1^\prime\big)
}{as symm algebra}
is the standard Diff($\mathbb{R}$) algebra of reparametrization symmetries at the AdS$_2$ boundary. 

Under the action of infinitesimal symmetry transformations generated by \eqref{residual gauge diffeos} with \eqref{constraint parameters}, the solution space transforms as
\eq{
    \delta_\xi X^+ =f\,\li X^+\ri^\prime-f^\prime\,X^+ \qquad\qquad \delta_\xi N =  \li f\,N\ri^\prime \qquad\qquad \delta_\xi J = f\,J^\prime \qquad\qquad  
    \delta_\xi M = f\,M^\prime \,. 
}{variations solution space}
In appendix \ref{sec:holography}, we give a holographic interpretation of these transformation laws. The identity
\eq{
    \delta_\xi\bigg( \frac1{X^+} \bigg)=\bigg( \frac{f}{X^+}\bigg)^\prime
}{variation zero mode} 
implies that the zero mode of $1/X^+$ does not transform and, therefore, can be set to some positive value, say, $1/\ell$ without constraining the asymptotic symmetries. This condition is required to obtain a well-defined variational principle for pure JT gravity \cite{Grumiller:2017qao}.

We conclude this subsection with an important observation: if we want to consider a leaky system where $J' \neq 0$ (see the discussion below the flux-balance law \eqref{mass change}), keeping the boundary source fixed on the phase space, $\delta J =0$, would generically break all the reparametrization symmetries at the boundary and enforce $f=0$, see \eqref{variations solution space}. As we shall argue in the next section, it is crucial to keep the reparametrization symmetries and allow variations of the sources, $\delta J \neq 0$, to deduce the flux-balance law \eqref{mass change} from the asymptotic charge algebra.

\subsection{Asymptotic charge algebra}

In this subsection, we discuss the variational principle and the charges associated with the asymptotic symmetries derived in the previous section. To do so, we use the covariant phase space methods that allow for a first-principle derivation of these charges \cite{Iyer:1994ys,Wald:1999wa,Barnich:2001jy} (see, e.g., \cite{Compere:2018aar,Ruzziconi:2019pzd} for reviews). The details of the computations of the symplectic structure and the associated charges for the current set-up are presented in detail in appendix \ref{sec:Derivation of the asymptotic charges}. 

As discussed there, the variation of the action \eqref{renormalized action} on solutions is given by a boundary term 
\begin{equation}
\label{variational principle}
    \delta S =-\int_{\partial \mathscr M} \!\!\!\!\extd t\,\Big[   \frac N{\ell^2}\,\delta J +\frac M{8\pi G} \,\delta\Big(\frac1{X^+}\Big)\Big]
\end{equation}
Reflective boundary conditions on the scalar field require either $\delta J=0=J^\prime$ while keeping $N$ fluctuating on the phase space (Dirichlet boundary conditions) or $N=0$ while keeping $J$ fluctuating (Neumann boundary conditions). As their names suggest, these conditions ensure the mass is conserved in time (see the flux-balance law \eqref{mass change}). In both cases, the variational principle is well-defined, i.e., $\delta S = 0$ on-shell. Indeed, the first term in \eqref{variational principle} is simply not there\footnote{%
This conclusion is immediate for Dirichlet boundary conditions and can be achieved for Neumann boundary conditions by adding a boundary term $-\int_{\partial\mathscr M}\extd x\sqrt{-h}\,\psi\,n^\mu\partial_\mu\psi=\frac{1}{\ell^2}\int_{\partial \mathscr M}\extd t\, N J$, see appendix \ref{sec:Derivation of the asymptotic charges} for our boundary conventions.}. The second term vanishes because the zero mode of $1/X^+$ is assumed to be fixed on the phase space (see the discussion below Equation \eqref{variation zero mode}).

In our case, we keep both $J$ and $N$ fluctuating on the phase space to consistently allow for leaks through the AdS boundary. The fact that the variational principle is not stationary on solutions should be expected since we are describing an open system. This is a standard feature appearing when considering leaky boundary conditions \cite{Compere:2020lrt,Fiorucci:2020xto,Ruzziconi:2020wrb}. Of course, if one added the bath whose dynamics is described by a variational principle $S_{\textrm{\tiny bath}}$ to close the system, we would expect to recover the usual statement $\delta S + \delta S_{\textrm{\tiny bath}} = 0$ as a consequence of the cancellation of the symplectic flux between what goes in and what goes out at the AdS boundary.
 
The expression of the co-dimension 2-form associated with the reparametrization symmetry $f$ is given by Equation \eqref{ktrdiffeo}. Using a field-dependent redefinition of the symmetry parameters (sometimes referred to as a change of slicing \cite{Barnich:2007bf,Adami:2020ugu,Ruzziconi:2020wrb}), 
\begin{equation}
    f=\tilde f \, X^+  
    \label{redef param}
\end{equation} 
and assuming $\delta \tilde f=0$, we find 
\begin{equation}
     k^{\rho t}_\xi [\phi; \delta \phi]\Big|_{\rho = 0}   = \delta \left( \tilde f \, \frac1{8\pi G} M \right)-\frac{\tilde f}{\ell^2}X^+\,N\delta J\,.
     \label{codimension 2}
\end{equation} 
This expression is a one-form on the phase space, which is integrable (namely $\delta$-exact) in the gravitational sector but remains non-integrable in the matter sector due to the leakiness \cite{Wald:1999wa,Barnich:2011mi}. The first term on the right-hand side of \eqref{codimension 2} can be readily integrated to give the gravitational charge
\begin{equation}
    Q_{\xi}^I[\phi] = \frac1{8\pi G}\,  \tilde f \, M \,.
\end{equation} 
In particular, if one sets $\tilde f=1$, this charge corresponds to the mass of the system. Taking the redefinition \eqref{redef param} into account, the symmetry algebra \eqref{as symm algebra} is replaced by an abelian algebra, 
\begin{equation}
    [\xi(\tilde f_1),\xi(\tilde f_2)]=0\,.
    \label{abelian symm}
\end{equation} 
Denoting the non-integrable piece in \eqref{codimension 2} by 
\begin{equation}
    F_{\xi} [\phi; \delta \phi] =-\frac{\tilde f}{\ell^2}X^+\,N\,\delta J
\end{equation} 
and using the Barnich--Troessaert bracket, which is adapted for non-integrable expressions \cite{Barnich:2011mi}, 
\begin{equation}
    \{ Q^I_{\xi_1}[\phi], Q^I_{\xi_2}[\phi] \}^* = \delta_{\xi_2}Q^I_{\xi_1}[\phi] + F_{\xi_2}[\phi;\delta_{\xi_1}\phi]
\label{modified bracket}
\end{equation} 
we obtain the charge algebra
\begin{equation}
    \{ Q^I_{\xi_1}[\phi], Q^I_{\xi_2}[\phi] \}^* = 0
    \label{charge algebra}
\end{equation}
forming a representation of the abelian symmetry algebra \eqref{abelian symm} at the boundary. 

From this charge algebra, we can derive a flux-balance law. The boundary-vector generating the reparametrization symmetries can be rewritten as $f \partial_t = \tilde{f} X^+ \partial_t = \tilde{f} \partial_{\tilde{t}}$, which defines a genuine time $\tilde t = \int^t\extd t' \frac{1}{X^+(t')}$. In terms of the genuine time, we have the flux-balance law
\eq{
   \frac{\extd}{\extd \tilde{t}}\,Q^I_{\xi}[\phi] = \partial_{\tilde t} Q^I_{\xi}[\phi] + \delta_{\xi(\tilde{f} = 1)}Q^I_{\xi}[\phi]
    =  Q^I_{\partial_{\tilde t}\xi}[\phi] -F_{\xi(\tilde{f} = 1)}[\phi;\delta_{\xi}\phi] 
}{flux-balance}
where in the second equality, we used \eqref{modified bracket} and \eqref{charge algebra} with $\xi_2 = \xi(\tilde f = 1)$ and $\xi_1 = \xi ( \tilde f)$. Therefore, keeping the sources fluctuating on the phase space, $\delta J \neq 0$, allows us to recover the flux-balance law controlling the exchanges between the leaky system we are considering and the environment. In particular, we recover the evolution of the mass discussed in the flux-balance law \eqref{mass change} as a particular case of \eqref{flux-balance} when setting $\tilde f = 1$.

\section{Leaky boundary conditions lead to diffeomorphism anomaly}\label{sec:lbcdiffeoanomaly}

In the previous section, we argued that a proper treatment of leaky boundary conditions for the scalar field requires keeping both the normalizable ($N$) and non-normalizable ($J$) modes turned on and as part of the phase space variables. In this section, we quantize this system on a fixed gravitational background using the path integral approach. As we shall argue, the standard diffeomorphism-invariant measure for the scalar field is ill-defined due to the interaction between normalizable and non-normalizable modes. This interaction is already responsible for the leaks, as evident from the right-hand sides of the flux-balance law \eqref{mass change} and \eqref{variational principle}. Therefore, all the subtleties discussed in this section are due to leakiness.\footnote{
We thank Andreas Karch and Lisa Randall for pointing out the relevance in higher dimensions of interaction terms between normalizable and non-normalizable modes.
}

\subsection{Problems of the naive approach}
\label{sec:Problems of the naive approach}

The one-loop partition function of the theory \eqref{renormalized action} is obtained by performing the path integral over the scalar field $\psi$ with fixed metric and dilaton background. In the Euclidean partition function 
\begin{equation}
    Z = \int D\psi DX  D g \, e^{-S[g_{\mu\nu},X, \psi]} \approx e^{-S_{\textrm{\tiny grav}}[\bar g_{\mu\nu}, \bar X]} \,Z_\psi
\end{equation} 
the quantity $S_{\textrm{\tiny grav}}$ denotes the two first terms of the action \eqref{renormalized action} evaluated on a background solution of the equations of motion. The wavy equality $\approx$ denotes that we used the saddle point approximation for the gravitational sector of the theory. The remaining factor to compute is the path integral of the scalar field on the fixed background (we drop the bar above the background solution and liberally integrate by parts from now on):
\begin{equation}
    Z_\psi = \int D \psi \, \exp \left[  - \frac{1}{2} \int_{\mathscr M}\!\! \extd^2 x\sqrt{g}\, \psi \Box \psi \right]
    \label{path integral}
\end{equation} 
with $\Box = g^{\mu\nu} \nabla_\mu \nabla_\nu$. In the standard case without exchange between AdS$_2$ and an external bath, $J$ is fixed and time-independent (usually set to zero). The partition function \eqref{path integral} can be computed by using the ultralocal definition of the path integral measure $D \psi$,
\begin{equation}
    1 = \int D \psi \, \exp \left[  - \frac{1}{2}\int_{\mathscr M}\!\! \extd^2 x\sqrt{g}\, \psi^2 \right]
    \label{diffeo invariance}
\end{equation} 
yielding the standard result $Z = (\det \Box)^{-\frac{1}{2}}$, see e.g.~\cite{Vassilevich:2003xt} on how to evaluate such functional determinants. The measure \eqref{diffeo invariance} is diffeomorphism invariant but not Weyl-invariant. Whenever a classical symmetry is not respected by the path integral measure it may become anomalous \cite{Fujikawa:1979ay}. In our context, the quantum theory acquires a Weyl anomaly, leading to a trace in the matter stress-energy tensor: $\langle {T^\mu}_{\mu} \rangle = -\frac{1}{24\pi}\,R$ (see, e.g., \cite{Grumiller:2002nm} for a derivation). 

In the leaky case depicted in Figure \ref{fig:leaks} and described in detail in section \ref{sec:Leaky boundary conditions}, the ``source'' $J$ is turned on at the boundary and is part of the phase space we have to quantize. As a result, the path integral measure \eqref{diffeo invariance} cannot be used because the falloffs \eqref{falloff conditions} imply
\begin{equation}
      \int_{\mathscr M}\!\! \extd^2 x  \sqrt{g}\, \psi^2 \sim \int\limits_0^\infty \extd\rho \int \extd t \, \Big(\frac{1}{\rho^2} + \mathcal{O}(1) \Big) \big( J + \rho N + \mathcal{O}(\rho^2) \big)^2  
      \sim \int\limits_0^\infty \extd\rho \int \extd t \, \Big(  \frac{1}{\rho^2} J^2 + \frac{2}{\rho} J  N + \mathcal{O}(1) \Big)
\label{divergences in the measure}
\end{equation} 
The two first terms in the expansion diverge, which means that the diffeomorphism-invariant path integral definition \eqref{diffeo invariance} cannot be taken as it is. Notice that this issue does not arise in the standard case discussed below \eqref{path integral} where $J$ is set to $0$. In the following, we present various attempts to circumvent this issue. We show that we cannot achieve simultaneously diffeomorphism invariance and locality of the path integral measure.

\subsection{Introduction of a cutoff}

A standard QFT technique of isolating divergences is introducing a cutoff $\epsilon\ll 1$, which in our case amounts to replacing the integral \eqref{divergences in the measure} by
\eq{
\int\limits_\epsilon^\infty \extd\rho \int \extd t \, \Big(  \frac{1}{\rho^2} J^2 + \frac{2}{\rho} JN + \mathcal{O}(1) \Big)=\frac{1}{\epsilon}\,\int\extd t\,J^2-2\ln\epsilon\,\int\extd t\,JN + {\cal O}(1)\,.
}{eq:cutoff1}
Employing this regularization, we can attempt to remedy the deficiency of the path integral measure \eqref{diffeo invariance} by treating the normalizable and non-normalizable modes as two different scalar fields $\psi_J=J(t)+{\cal O}(\rho^2)$ and $\psi_N=\rho\,N(t)+{\cal O}(\rho^2)$. We define the regularized path integral measure\footnote{
We thank Hao Geng for discussions on this regularized measure.
}
\eq{
 1 = \int D \psi_J\, D\psi_N \, \exp \left[  - \frac{1}{2}(\psi_J,\,\psi_N)_A\, \mat_{AB}\, (\psi_J,\,\psi_N)_B \right]
}{eq:cutoff2}
where the matrix $\mat_{AB}$ according to \eqref{eq:cutoff1} is given by
\eq{
\mat=\begin{pmatrix}
\frac{1}{\epsilon} + {\cal O}(1) & -\ln\epsilon + {\cal O}(1) \\
-\ln\epsilon + {\cal O}(1) & {\cal O}(1)
\end{pmatrix}\,.
}{eq:cutoff3}
Diagonalizing the matrix \eqref{eq:cutoff3} yields the eigenvalues $\lambda_1={\cal O}(1)$, $\lambda_2=\frac1\epsilon+{\cal O}(1)$ associated with the eigenvectors 
\eq{
\psi_1=\psi_N+{\cal O}(\epsilon\,\ln\epsilon)\qquad\qquad\psi_2=-\frac{\psi_J}{\epsilon\,(\ln\epsilon+{\cal O}(1))} + {\cal O}(1)\,.
}{eq:cutoff4}

The cutoff $\epsilon$ breaks diffeomorphism invariance, so we need to remove it eventually if we desire a diffeomorphism invariant quantization scheme. If all divergences were monomial, we might rescale our measure suitably by some overall power of $\epsilon$ to render the rescaled measure finite, see the discussion in section 5.1 of \cite{Geng:2023ynk}.

However, the second eigenvector in \eqref{eq:cutoff4} depends logarithmically on the cutoff to leading order, so we cannot use this procedure here. Since logarithmic behavior on the cutoff often signifies the presence of an anomaly (e.g., the anomalous leading order divergent contribution to entanglement entropy in a CFT$_2$ \cite{Holzhey:1994we, Calabrese:2004eu}), we take the results above as an indication that (at least some of the) diffeomorphisms become anomalous in the leaky quantum theory. For this conclusion, leakiness was essential. Indeed, if we only had the normalizable mode $\psi_N$ (or the non-normalizable mode $\psi_J$, but not both simultaneously), we could define a (regularized) diffeomorphism invariant path integral measure (and there would be no logarithmic dependence on the cutoff $\epsilon$). The subtlety highlighted above is due to the interplay between $\psi_N$ and $\psi_J$. In the next subsections, we address this fundamental issue from various perspectives.

\subsection{Non-local measure or diffeomorphism anomaly}

As described above, the issue in the path integration arises because of the interplay between normalizable and non-normalizable modes. Hence, a natural way to proceed is to treat normalizable and non-normalizable modes separately in the path integral, but avoiding the issues with the diagonalization explained in the previous subsection. To do so, we consider a non-normalizable solution $\psi_J(t,\rho)$,
\begin{equation}
   \Box \psi_J(t ,\rho) = 0\qquad \qquad \lim_{\rho\to 0} \psi_J (t ,\rho)  \to J(t )
\end{equation} 
so that
\begin{equation}
    \psi (t ,\rho) - \psi_J (t ,\rho) \to \mathcal O (\rho) \qquad \text{when} \qquad {\rho \to 0}\,.
    \label{falloffs after substraction}
\end{equation} 
As a first step, we perform the path integral \eqref{path integral} with fixed sources $J(t)$ using the diffeomorphism invariant measure
\begin{equation}
     1 = \int D \psi \, \exp \left[  - \frac{1}{2} \int \extd^2 x \sqrt{g}\,\big(\psi - \psi_J\big)^2 \right]
\end{equation} 
which is perfectly well-defined thanks to the falloffs \eqref{falloffs after substraction}. This leads to 
\begin{equation}
    Z[J] = \exp\left[  - \frac{1}{2} \int \extd t  \extd t'  \sqrt{h(t)} \sqrt{h(t')}\, J(t) J(t') G(t - t') \right] (\det \Box )^{-\frac{1}{2}}
\end{equation} 
where $h(t)$ is the boundary metric and $G(t - t') \sim 1/(t-t')^2$ is the boundary $2$-point correlation function \cite{Witten:1998qj}. At this stage, the procedure is completely diffeomorphism invariant. In the second step, to capture interactions with the external bath, one has to integrate over the external sources. 

Several choices are possible at this stage. If one wants to preserve bulk diffeomorphism invariance, the measure for the sources\footnote{%
We thank Suvrat Raju for proposing this measure.} 
    \begin{equation}
        1 = \int DJ\, \left[  - \frac{1}{2} \int \extd t  \extd t' \sqrt{h(t)} \sqrt{h(t')}\, J(t) J(t') G(t - t') \right]
    \end{equation} 
is not local but bi-local. For the free scalar field, this measure makes the sources trivial in the theory. In the presence of interactions, non-trivial features might appear. We comment on closing the system by adding bath degrees of freedom in the concluding section. 

Instead, if one wants to preserve locality in the definition of the measure, then one necessarily breaks parts of the bulk diffeomorphisms. Indeed, the local measure
    \begin{equation}
        1 = \int DJ\, \left[  - \frac{1}{2} \int \extd t \sqrt{h(t)}\, J(t)^2 \right] \label{def measure boundary diffeos}
    \end{equation} 
is boundary diffeomorphism invariant but not boundary Weyl-invariant. This implies that the radial bulk diffeomorphisms (inducing boundary Weyl rescalings) are broken. Let us have a closer look at these symmetries in the Fefferman--Graham gauge \eqref{FGgauge}. The generators of the residual gauge diffeomorphisms are given in \eqref{residual gauge diffeos} where we recall that $f(t)$ and $\sigma (t)$ induce boundary diffeomorphisms and boundary Weyl rescalings, respectively, $\delta_\xi \sqrt{h} = (f \sqrt{h})' - \sigma \sqrt{h}$ and $\delta_\xi J = f J'$. One can easily check that the measure $DJ$ in \eqref{def measure boundary diffeos} is preserved under the action of diffeomorphisms \eqref{residual gauge diffeos} generated by $f(t)$, but not under those generated by $\sigma (t)$.
    
Alternatively, the local measure
    \begin{equation}
        1 = \int DJ\, \left[  - \frac{1}{2} \int \extd t \, J(t)^2 \right]
        \label{def measure boundary weyl}
    \end{equation} 
is boundary Weyl-invariant, but not boundary diffeomorphism invariant. Indeed, the radial bulk diffeomorphisms (those generated by $\sigma(t)$ in \eqref{residual gauge diffeos}) preserve the measure \eqref{def measure boundary weyl}, while the tangent bulk diffeomorphisms (those generated by $f(t)$ in \eqref{residual gauge diffeos}) do not preserve the measure. 
    
Since it is impossible to have a local measure that preserves both boundary Weyl and diffeomorphism invariance, locality together with finiteness of the measure imply a breaking of bulk diffeomorphism invariance in the quantum theory at one loop. A loophole in this argument is that we could use the dilaton to make the measure finite, while keeping diffeomorphism invariance. This possibility is pursued in appendix \ref{sec:6}, but does not lead to a conserved stress-energy tensor either, and moreover does not seem natural for minimally coupled scalar fields.

\section{Local and Weyl-invariant measure}
\label{sec:Local and Weyl invariant measure}

In the previous section, we have seen that locality and bulk diffeomorphism invariance cannot be simultaneously preserved at one loop in the presence of leaky boundary conditions. We now investigate a local path integral measure that is not bulk diffeomorphism invariant but bulk Weyl-invariant \cite{Karakhanian:1994gs,Jackiw:1995qh}. 

The measure 
\begin{equation}
     1 = \int D \psi \, \exp \left[  - \frac{1}{2}\int \extd^2 x \, \psi^2 \right]
     \label{Weyl measure}
\end{equation} 
is often considered as an alternative of \eqref{diffeo invariance} and has the advantage of being well-defined in the presence of leaky boundary conditions. This measure does not break all the bulk diffeomorphisms: The area-preserving diffeomorphisms (by which we mean $\partial_\mu \xi^\mu = 0$) are still symmetries of \eqref{Weyl measure}. 

Before discussing the consequences of using the Weyl-invariant measure \eqref{Weyl measure} on the one-loop path integral, let us first review some aspects of the standard diffeomorphism invariant measure \eqref{diffeo invariance}. In the latter case, the effective action is given by the Polyakov action 
\begin{equation}
    \Gamma_{\text{eff}}^D[g] = - \frac{c}{96\pi}\int \extd^2x \sqrt{g}\, R \Box_g^{-1} R 
    \label{Polyakov}
\end{equation}  
with unity central charge, $c=1$, for one massless scalar field, and $\sqrt{g}\,\Box_g(x)=\int\extd^2y \delta^2(x-y)\partial_\mu \sqrt{g}\,g^{\mu\nu}\partial_\nu$. 

To render the effective action local, one can introduce an \textit{ancillary field}\footnote{%
An auxiliary field has equations of motion that can be solved algebraically and re-injected into the action. Here, we use the term ``ancillary field'' for $\chi$ as its equations of motion are not algebraic, but its role is somehow analogous to the one of an auxiliary field.} 
$\chi$ to obtain
\begin{equation}
    \Gamma_{\text{eff}}^D[g,\chi] = - \frac{c}{24 \pi} \int \extd^2 x \sqrt{g}\, \big[g^{\mu\nu}(\partial_\mu \chi) (\partial_\nu \chi) + \chi R\big]
    \label{auxiliary action}
\end{equation} 
The $\chi$-equation of motion $\Box_g \chi = \frac12 R$ can be re-injected into the action \eqref{auxiliary action} to recover \eqref{Polyakov}. The stress-energy tensor is 
\eq{
 \langle T_{\mu\nu}^D \rangle = \frac{2}{\sqrt{g}}\frac{\delta \Gamma^{D}_{\text{eff}}[g,\chi]}{\delta g^{\mu\nu}}= -\frac{c}{12\pi}\left((\nabla_\mu \chi)(\nabla_\nu\chi)-\frac12 g_{\mu\nu} (\nabla \chi)^2-\nabla_\mu\nabla_\nu\chi +\Box_g \chi g_{\mu\nu} \right)\,.
}{eq:nolabel}
From this expression, one can now explicitly check that the quantum theory is diffeomorphism invariant but not Weyl-invariant,
\begin{equation}
    g^{\mu\lambda}\nabla_\lambda \langle T^D_{\mu\nu} \rangle =0 \qquad \qquad  g^{\mu\nu}\langle T^D_{\mu\nu}\rangle   =-\frac c{24\pi}R\,.
    \label{stress tensor diffeos}
\end{equation}  

Let us now repeat the same steps for the Weyl-invariant path integral measure \eqref{Weyl measure}. In this case, the effective action 
\begin{equation}
    \Gamma^W_{\text{eff}} [g] =\frac{c}{96\pi} \int \extd^2 x \,  \mathcal{R} \big(\sqrt{g}\, \Box_g\big)^{-1} \mathcal{R}   \label{Weyleffectiveaction}
\end{equation} 
 depends on the Ricci scalar $\mathcal{R}$ associated with the rescaled metric $\gamma_{\mu\nu} = \frac{g_{\mu\nu}}{\sqrt{g}}$ and $\gamma^{\mu\nu} = \sqrt{g} g^{\mu\nu}$. Note that $\mathcal{R}$ is related to the Ricci scalar associated with $g_{\mu\nu}$ through $R = \frac1{\sqrt{g}} \li   \mathcal  R - \Box_{\gamma} \ln \sqrt{g} \ri$ where $\Box_\gamma=\sqrt{g}\Box_g$. As in the standard case, to render the effective action local, one can introduce an ancillary field $\chi$ to obtain
\begin{equation}
\Gamma_{\text{eff}}^W[\gamma,\chi] = - \frac{c}{24 \pi} \int \extd^2 x \, \big[\gamma^{\mu\nu}(\partial_\mu \chi) (\partial_\nu \chi) + \chi \mathcal R\big]\,.
    \label{Weyl auxiliary action}
\end{equation} 
Again, the $\chi$-equation of motion $\Box_\gamma\chi = \frac12 \mathcal R$ can be re-injected into the action \eqref{Weyl auxiliary action} to recover \eqref{Weyleffectiveaction}.

The stress-energy tensor 
\begin{equation}
   T^W_{\mu\nu} =\frac{2}{\sqrt{g}}\frac{\delta \Gamma^{W}_{\text{eff}}[g,\chi]}{\delta g^{\mu\nu}} = -\frac{c}{12\pi}\left((\partial_\mu \chi)(\partial_\nu\chi)-\frac12 \gamma_{\mu\nu} (\partial \chi)^2-D_\mu D_\nu\chi + {\frac12}\gamma_{\mu\nu}\Box_\gamma \chi  \right)\label{eq:T}
\end{equation}
contains $D_\mu$, the covariant derivative with respect to $\gamma_{\mu\nu}$.\footnote{%
See appendix \ref{app:IyerWaldeffectiveaction} for explicit expressions of \eqref{Weyleffectiveaction} and \eqref{eq:T} in terms of the metric $g$.} 
Setting $\chi$ on-shell, we have 
\begin{equation}
\boxed{\phantom{\bigg(}
       \gamma^{\mu\lambda}D_{\lambda}\,\langle T^W_{\mu\nu}\rangle =\frac{c}{48\pi}\,\partial_\nu \mathcal R \qquad\qquad \gamma^{\mu\nu}\,\langle T^W_{\mu\nu}\rangle=0\,.
       \phantom{\bigg)}}
     \label{stress tensor Weyl}
\end{equation} 
Since $T^W_{\mu\nu}$ is a symmetric traceless tensor, we have $\sqrt{g}g^{\mu\rho}\nabla_\rho T^W_{\mu\nu}= \gamma^{\mu\rho}D_{\rho}T^W_{\mu\nu}$. Comparing \eqref{stress tensor Weyl} with \eqref{stress tensor diffeos}, we see that the anomaly has been moved from the Weyl symmetry to the diffeomorphisms. In other words, the effective action \eqref{Weyl auxiliary action} is Weyl-invariant, but not diffeomorphism invariant.

The Ricci scalar $\mathcal R$ constructed from the rescaled metric $\gamma$, and hence the anomaly \eqref{stress tensor Weyl}, depend on the choice of coordinates. For instance, in coordinates where $\sqrt{g}=1$ the anomaly expression $\partial_\nu\mathcal R$ vanishes. However, for self-consistency we should use the set of coordinates \eqref{FGgauge} that defined our boundary conditions. For our specific case, we have
\eq{
\mathcal R =  \frac1{(X^+)^2}\,M-\frac{\ell^2}2\frac{((X^+)')^2}{(X^+)^2}+\ell^2 \frac{(X^+)''}{X^+}+ {\cal O}(\rho^2)
= e^{-2\Phi}\,M+2{\cal L}_{\textrm{\tiny tS}} + {\cal O}(\rho^2) 
}{eq:calR}
where, in the second equality, we have used the notation of appendix \ref{sec:holography}. Thus, the Ricci scalar $\mathcal R$ to leading order is given by the subleading metric coefficient $g_{tt}$ \eqref{falloff conditionsmetric} that contains state-dependent information. 

Neglecting backreactions, for a JT black hole background \eqref{eq:JTsol} we get $\mathcal R=M=\rm const.$ and hence the anomaly in \eqref{stress tensor Weyl} vanishes. However, genuine leaky boundary conditions at the quantum level require including backreactions so that the mass $M$ can actually decrease in time and the black hole can evaporate. In that case, the diffeomorphism anomaly discussed in this section will have a direct physical effect. In the next section, we discuss the impact of the anomaly on the gravitational Gauss law.

\section{Consequences for the Gauss law}\label{sec:5}

As discussed above, if one wants to preserve a local definition for the path integral measure, one has to break part of the bulk diffeomorphisms. The remaining subset of diffeomorphisms that are symmetries of the theory depends on the specific choice of measure. In this section, we discuss the implication of violating diffeomorphism invariance of the theory at one loop when computing the asymptotic charges and derive an explicit violation of the Gauss law.

\subsection{Classical Noether identities}

Let us first review the standard relation between diffeomorphism invariance and Noether identities for the classical theory \eqref{renormalized action}. To do so, we use the covariant phase space framework discussed in \cite{Barnich:2001jy, Compere:2007az} (see also \cite{Compere:2018aar,Ruzziconi:2019pzd} for reviews). It is convenient to introduce the following notations: $\phi = (g_{\mu\nu}, X, \psi)$ and $L[\phi] = L_{\textrm{\tiny JT}}[g,X] + L_m[\psi, g]$, with $L_{\textrm{\tiny JT}}[g,X] = \frac{\sqrt{g}}{16 \pi G} X [R- 2 \Lambda]$ and $L_m[\psi, g] = \frac{1}{2} \sqrt{g} \,g^{\mu\nu}(\partial_\mu \psi)(\partial_\nu \psi)$. Setting the Euler--Lagrange derivatives of $L$ with respect to all fields,
\begin{align}
\frac{\delta L}{\delta g^{\mu\nu}}&= \frac{\sqrt{g}}{16\pi G} \left(\nabla^\mu \nabla^\nu X - g^{\mu\nu} \nabla^2 X - g^{\mu\nu} \Lambda X\right)  -\frac{\sqrt{g}}2\, T^{\mu\nu} \\
 \frac{\delta L}{\delta X}&=\frac{\sqrt{g}}{16\pi G} \left( R - 2 \Lambda\right)  \\  \frac{\delta L}{\delta \psi}&=-\sqrt{g}\,\nabla^2 \psi 
\label{EOM matter2}
\end{align} 
to zero recovers the equations of motion \eqref{EOM matter}. In the identity
\begin{equation}
    \delta_\xi \phi\, \frac{\delta L}{\delta \phi^i} = \partial_\mu S^\mu_\xi [\phi] + \xi^\mu N_\mu [\phi]
    \label{step 1}
\end{equation} 
the Lie derivative, $\delta_\xi \phi^i \equiv \mathcal{L}_\xi \phi^i$, of the field $\phi^i$ with respect to $\xi$ multiplies the Euler--Lagrange derivative of $L$ with respect to $\phi^i$. The right-hand side of \eqref{step 1} is obtained by the Leibniz rule to remove the derivative on the diffeomorphism parameters $\xi^\mu$ and keeping the total derivative term defining the weakly vanishing Noether current $S^\mu_\xi[\phi]$ given by
\begin{equation}
    S^\mu_\xi[\phi] = 2\xi_\nu\,\frac{\delta L}{\delta g_{\mu \nu}}\,.               
    \label{weakly vanishing Noether current}
\end{equation} 
The second term on the right-hand side of \eqref{step 1} exhibits the Noether identities $N_\mu$, which identically vanish off-shell as a consequence of diffeomorphism invariance. Indeed, one can check explicitly 
\begin{equation}
         N_\mu [\phi]= \frac{\delta L_{\textrm{\tiny JT}}}{\delta X}\,\partial_\mu X + \frac{\delta L_m}{\delta \psi}\,\partial_\mu \psi - 2 g_{\mu\alpha} \nabla_\nu \left(\frac{\delta L}{\delta g_{\nu\alpha}}\right) = \frac{\delta L_m}{\delta\psi}\,\partial_\mu\psi + \sqrt{g}\,\nabla^\nu T_{\mu\nu}=0
\end{equation} 
with the stress-energy tensor \eqref{eq:angelinajolie}. The on-shell conservation of the stress-energy tensor \eqref{eq:conserved} is a direct consequence of the Noether identities. 

\subsection{Quantum gravitational Gauss law}

Semi-classically, we consider the theory $L[\phi] = L_{\textrm{\tiny JT}}[g,X]+ L_{\text{eff}}[g,\chi]$ where $L_{\text{eff}}[g]$ is the effective Lagrangian obtained after path integration over the matter field $\psi$. As discussed in section \ref{sec:Local and Weyl invariant measure}, to make the effective action local and be able to use our methods, we introduce an ancillary field $\chi$ in the effective action. The precise form of $L_{\text{eff}}[g, \chi]$ depends on the particular measure prescription for the integration over the matter field. For instance, if the diffeomorphism invariant measure \eqref{path integral} is chosen, then $L_{\text{eff}}[g, \chi]$ is the Polyakov action \eqref{auxiliary action}. Alternatively, if \eqref{Weyl measure} is chosen, then $L_{\text{eff}}[g, \chi]$ is the Weyl invariant action \eqref{Weyl auxiliary action}. The expectation value of the stress-energy tensor is obtained from
\begin{equation}
    \langle T_{\mu\nu} \rangle = \frac{2}{\sqrt{g}}\frac{\delta L_{\text{eff}}[g]}{\delta g^{\mu\nu}}\,.
\end{equation} 
The weakly vanishing Noether current has the same expression as in \eqref{weakly vanishing Noether current} but with $L_m$ replaced by $L_{\text{eff}}$. The Noether identities 
\begin{equation}
         N_\mu [\phi]=  \frac{\delta L_{\textrm{\tiny JT}}}{\delta X}\,\partial_\mu X -2 g_{\mu\alpha} \nabla_\nu \frac{\delta L}{\delta g_{\nu\alpha}}+  \frac{\delta L_{\text{eff}}}{\delta \chi}\,\partial_\mu \chi
        =  \sqrt{g}\, \nabla^\nu \langle T_{\mu\nu} \rangle +  \frac{\delta L_{\text{eff}}}{\delta \chi}\,\partial_\mu \chi
\end{equation} 
vanish only for quantum theories preserving diffeomorphism invariance. From now on, we focus on the path integral measure \eqref{Weyl measure} giving rise to the effective action \eqref{Weyl auxiliary action} for the matter sector. The violation in the Noether identities can be deduced from \eqref{eq:T} and reads explicitly
\begin{equation}
    N_\mu [\phi] = \frac{c}{24\pi}\,\partial_\mu\big(\sqrt{g}\, \Box_g\chi\big)\, .
\end{equation} 
We continue the analysis keeping this contribution. More precisely, we show that the standard statement that gravitational charges reduce to surface charges (\textit{Gauss law}) does not hold if this contribution is non-vanishing. More details about the standard derivation without this contribution can be found in \cite{Barnich:2001jy, Compere:2007az}. The Barnich--Brandt co-dimension 2 form $\mathbf{k}_\xi$ and the presymplectic form $\mathbf{W}$ are defined as
\begin{equation}
    \mathbf{k}_\xi [\phi] =- I_{\delta \phi}^{n-1}  \mathbf{S}_\xi[\phi]\qquad  \qquad \mathbf{W}[\phi; \delta \phi, \delta \phi] = \frac{1}{2} I^n_{\delta \phi} \left( \delta \phi^i \frac{\delta \mathbf{L}}{\delta \phi^i} \right)
\end{equation} with $\mathbf{k}_\xi  = k^{\mu\nu}_\xi (\extd^0x)_{\mu\nu}$, $\mathbf{S}_\xi = S^\mu_\xi (\extd^1x)_\mu$, $\mathbf{W}_\xi = W^\mu_\xi (\extd^1x)_\mu$ and $\mathbf{L} = L (\extd^2x)$. Here $I_{\delta \phi}^{k}$ denotes the Anderson homotopy operator acting on $k$-forms. Using the properties of this operator, we have
\begin{multline}
    \extd\mathbf{k}_\xi[\phi] = -\extd I_{\delta \phi}^{n-1}  \mathbf{S}_\xi[\phi] = \delta \mathbf{S}_\xi[\phi] - I^n_{\delta \phi} \extd\mathbf{S}_\xi[\phi]
    \approx - I^n_{\delta \phi} \extd\mathbf{S}_\xi[\phi] \\
    = - I^n_{\delta \phi} \left( \delta_\xi \phi^i \frac{\delta \mathbf{L}}{\delta \phi^i}  \right)  - I^n_{\delta \phi} (\xi^\mu N_\mu [\phi] (\extd^2x)) = \mathbf{W}[\phi; \delta_\xi \phi, \delta \phi]  - I^n_{\delta \phi} (\xi^\mu N_\mu [\phi](\extd^2x))
   \label{Gauss law violation}
\end{multline} 
where $\approx$ means that we have used the equations of motion by taking into account the backreaction. In Lorentzian signature, integrating over a Cauchy slice $\Sigma$, we obtain
\begin{equation}
\label{breaking Gauss law}
   \boxed{\phantom{\Big(} 
   \underbrace{\int_{\Sigma} \mathbf{W}[\phi; \delta_\xi \phi, \delta \phi]}_{\text{Contraction of presymplectic form}} = \underbrace{\int_{\partial \Sigma}  \mathbf{k}_\xi[\phi]}_{\text{Surface charge}} + \underbrace{\int_{\Sigma}  I^n_{\delta \phi} \left[\xi^\mu N_\mu[\phi] (\extd^2x)\right]}_{\text{Gauss law breaking}}\,. \phantom{\Big)}}
\end{equation} 
This result states that the gravitational charges do not reduce to simple boundary terms. The breaking term on the right-hand side is due to the violation of the Noether identities. In appendix \ref{app:IyerWaldeffectiveaction}, we provide a more explicit expression  for the breaking term (see Equation \eqref{expression breaking gauss law}) and re-derive this result using the Iyer--Wald approach \cite{Iyer:1994ys}, following the steps presented in \cite{McNees:2023tus}. 

The breaking of the standard gravitational Gauss law that we find in \eqref{breaking Gauss law} is reminiscent of what was observed in higher dimensions, where coupling AdS to an external bath breaks the Gauss law and is interpreted as gravitons becoming massive through a Higgs mechanism \cite{Geng:2020qvw, Geng:2021hlu}. However, as stated in the introduction, this interpretation no longer holds in 2d gravity since there are no propagating degrees of freedom.

\section{Discussion and generalization}\label{sec:7}

In this concluding section, we address implications for black hole evaporation and the addition of a bath system coupled to the black hole. We also comment on some generalizations of the observations raised in this paper. 

\subsection{Implications for black hole evaporation and bath systems}

Our starting point was to allow leakiness at the AdS boundary so that (large) black holes in AdS can evaporate. Classically, leakiness implies a non-trivial flux-balance law \eqref{mass change} that relates the time derivative of the mass function to (scalar) news. A pivotal technical aspect was that we considered the non-normalizable mode of the scalar as state-dependent, in contrast to a standard AdS/CFT setup where these modes are treated instead as sources that do not vary over the state space. We explained our reasoning for this choice in the last paragraph of subsection \ref{sec:ASA}.

Semi-classically, we have seen that the leakiness comes with some quantum baggage: The path integral measure was either ill-defined, non-local, or led to a non-conserved expectation value of the stress-energy tensor. We pursued the last option in detail and showed how the diffeomorphism anomaly \eqref{stress tensor Weyl} leads to a breaking of the Gauss law, reminiscent of what happens in higher dimensions \cite{Geng:2020qvw, Geng:2021hlu}.

These conclusions could potentially be circumvented by adding the bath, which would close our open quantum system. While details of the full system depend on the precise model of the bath, one should always be able to eliminate the diffeomorphism anomaly. For the specific model we studied --- JT gravity with a minimally coupled massless scalar field, with both normalizable and non-normalizable modes switched on --- such a resolution might be pursued along the lines of \cite{Geng:2023ynk}. The technical key point of that work is a diagonalization of the two scalar modes into a normalizable and a non-normalizable mode, both of which have support on the gravity \emph{and} the bath sides. By contrast, just working on the AdS side, it was impossible to diagonalize the scalar modes, and the off-diagonal interaction term between normalizable and non-normalizable modes was the origin of the breaking of the Gauss law. We expect that the bath details do not matter for the overall conclusion that adding a bath closes the quantum system and should restore diffeomorphism invariance semi-classically.

\subsection{Higher dimensions}

Let us generalize the discussion of section \ref{sec:Problems of the naive approach} for Einstein gravity in even\footnote{%
In odd dimensions, a similar argument works, but the expansion \eqref{eq:wtf} acquires an additional factor $\ln\rho$ in front of $N$.} 
dimension $d+1$ ($d>1$). We consider gravity coupled to a massless scalar field $\psi$ with the bulk action
\begin{equation}
    S[g_{\mu\nu},\psi] = \frac{1}{16 \pi G} \int_{\mathscr M} \extd^{d+1}x \sqrt{-g}\,\big(R- 2 \Lambda\big) + \frac{1}{2} \int_{\mathscr M} \extd^{d+1}x \sqrt{-g}\, g^{\mu\nu} \partial_\mu \psi \partial_\nu \psi 
\end{equation} 
where $\Lambda = - \frac{d(d-1)}{2\ell^2} < 0$. The matter sector of the theory is not Weyl-invariant in dimension $d>1$ (to restore such an invariance, one would need to add the interaction term $\frac{d-1}{8d}\,\psi^2\,R$ to the action). On-shell, in Fefferman--Graham gauge we have the falloffs
\eq{
    g_{\rho\rho} = \frac{\ell^2}{\rho^2}\,, \qquad g_{ab}= \frac{1}{\rho^2}\bar{g}_{ab} +  \mathcal{O}(\rho^0)\,, \qquad g_{\rho a} = 0\,, \qquad \psi = J + \ldots + \rho^d N + \mathcal{O}(\rho^{d+1})
}{eq:wtf} 
where $\bar{g}_{ab}$ is a fixed boundary metric, $N$ and $J$ are, respectively, the normalizable and non-normalizable modes of the scalar field. To integrate out the matter field in the path integral, we have to make a choice of measure. The natural diffeomorphism-invariant measure $D\psi$ for the scalar field is defined through
\begin{equation}
    1 = \int D\psi\, \exp\Big[ - \int \extd^{d+1} x \sqrt{-g}\, \psi^2 \Big]\,.
    \label{path integral measure in d dimensions}
\end{equation} 
If we turn off the non-normalizable $J$ mode, we get 
\begin{equation}  
    \int \extd^{d+1} x \sqrt{-g}\, \psi^2 \sim \int \extd^{d} x\, \rho^{d-1} N^2+ \ldots 
\end{equation} 
which is finite at small $\rho$. In this case, the usual normalization \eqref{path integral measure in d dimensions} can be used. However, this normalization cannot be achieved if the non-normalizable mode $J$ is turned on as well. Indeed, noticing that $\sqrt{-g} = \rho^{-(d+1)} \ell \sqrt{-\bar{g}}$, we have 
\begin{equation}  
    \int \extd^{d+1} x \sqrt{-g}\, \psi^2 \sim \lim_{\epsilon\to 0}\int\limits_\epsilon^\infty \extd \rho \int \extd^{d}x\, \left(\frac{1}{\rho^{d+1}} J^2+ \ldots + \frac{1}{\rho} J N + \ldots\right)
\end{equation} 
which is divergent and therefore incompatible with \eqref{path integral measure in d dimensions}. As in the 2d case, since $J$ is dynamical, it cannot be substracted from the path integral normalization. Therefore, also in higher dimensions one has to use another definition of the measure that will either be non-local or break part of the diffeomorphisms and, consequently, the Gauss law.

\subsection{de Sitter}

So far, we have considered leaky boundary conditions in asymptotically AdS spacetimes, which are suitable to describe interactions between the gravitational system and an external bath. While this set-up is a bit unconventional in AdS, leaky boundary conditions are more common for other types asymptotics. We comment on the possible extension of our results for de Sitter (dS) spacetimes. 

\begin{figure}[htb]
    \centering
\begin{tikzpicture}[scale=0.5]
	\tiny
\draw[red,opacity=0] (3.5,-9.1) -- (12.5,-9.1) -- (12.5,1) -- (3.5,1) -- cycle;
\coordinate (P) at (5,-7);
\coordinate (Q) at (11,-7);
\coordinate (R) at (11,-1);
\coordinate (S) at (5,-1);
\draw (P) -- (Q) -- (R) -- (S) -- cycle;
\draw ($(P)!0.8!(Q)$) node[below,outer sep=4pt]{$\mathscr I^-_{\text{dS}}$};
\draw ($(R)!0.8!(S)$) node[above,outer sep=4pt]{$\mathscr I^+_{\text{dS}}$};
\draw ($(P)!0.5!(S)$) node[above,rotate=90,outer sep=3pt]{North pole};
\draw ($(Q)!0.5!(R)$) node[above,rotate=-90,outer sep=3pt]{South pole};
\path [blue] (5,-3) edge[bend left=30] (8,-3) -- (8,-3) edge[bend right=30] (11,-3);
\draw (9.5,-4) node[blue]{$\Sigma$};
\draw[black!50,densely dashed] (P) -- (R);
\path [-{Latex[width=1mm]},draw=red] (6,-3.5) -- (10,0.5);
\path [-{Latex[width=1mm]},draw=red] (6.25,-3.75) -- (10.25,0.25);
\path [-{Latex[width=1mm]},draw=red] (6.5,-4) -- (10.5,0);
\end{tikzpicture}
\caption{dS spacetime is leaky}
    \label{fig:dS}
\end{figure}
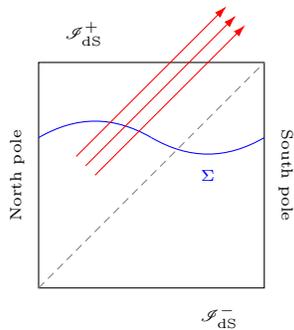

In dS spacetime, imposing boundary conditions at the future spacelike boundary $\mathscr{I}^+_{\text{dS}}$ (see Figure \ref{fig:dS}) is a delicate issue \cite{Anninos:2011jp,Anninos:2010zf,Ashtekar:2019khv,Ashtekar:2014zfa,Compere:2019bua,Compere:2020lrt,Compere:2023ktn} since it can drastically
restrict the Cauchy problem. Indeed, if one prescribes generic data on a Cauchy slice $\Sigma$ in the bulk, one naturally ends up with leaky boundary conditions at the future spacelike boundary $\mathscr{I}^+_{\text{dS}}$. Leaky boundary conditions are therefore appealing in this context.


Most of the computations performed in this paper could easily be adapted to the dS case, just by changing the sign of the cosmological constant ($\ell^2 \to - \ell^2$) and considering an expansion near the future boundary $\mathscr{I}^+_{\text{dS}}$ ($\rho$ now becomes a timelike coordinate, while $t$ becomes spacelike). In particular, the problem in the definition of the measure discussed in section \ref{sec:Problems of the naive approach} still applies there and the non-normalizable modes cannot be turned off if one wants to preserve a generic dynamics. It would be interesting to investigate the potential consequences of our work on cosmology and inflation where semi-classical computations play an important role (see \cite{Mukhanov:1990me,Mukhanov:2007zz} and references therein). 

In conclusion, the subtle interplay between leakiness and 1-loop effects uncovered in this paper is not tied to the specific model we studied in detail --- JT gravity with minimally coupled massless matter --- but appears to be a rather generic feature of semi-classical gravity with leaky boundary conditions.


\paragraph{Acknowledgements} We warmly thank Hao Geng, Andreas Karch, Suvrat Raju and Lisa Randall for insightful discussions and collaboration at early stages of the project. We also thank Jan de Boer, Adrien Fiorucci, Laurent Freidel, Diego Hofman, Thomas Mertens, Dominik Neuenfeld and Dima Vassilevich for useful discussions. DG~thanks CZ~for hosting his research stay at the Perimeter Institute during the final stage of this project.

DG was supported by the Austrian Science Fund (FWF), projects P 32581, P 33789, and P 36619. Part of this research was conducted while DG was visiting the Okinawa Institute of Science and Technology (OIST) through the Theoretical Sciences Visiting Program (TSVP). 
RR is supported by the Titchmarsh Research Fellowship at the Mathematical Institute and by the Walker Early Career Fellowship in Mathematical Physics at Balliol College. 
Research at Perimeter Institute is supported in part by the Government of Canada through the Department of Innovation, Science and Economic Development Canada and by the Province of Ontario through the Ministry of Colleges and Universities. 


\appendix

\section{Derivation of asymptotic charges}
\label{sec:Derivation of the asymptotic charges}

In this appendix, we provide a detailed derivation of the asymptotic charges for the model of interest discussed in section \ref{sec:Leaky boundary conditions}. We derive the symplectic potential and symplectic current for \eqref{renormalized action} and then compute the charges. For a review on the computation of charges in 2d gravity, see e.g. \cite{Grumiller:2021cwg}.

First, we consider the variational principle for JT gravity without matter. It has been studied in detail in \cite{Grumiller:2017qao} (see Equation (4.1) of that reference). Denoting summarily all the fields by $\phi$, the holographically renormalized JT action 
\begin{multline}
S_{\textrm{\tiny JT}}[\phi]=\int_{M}\!\! \extd^2xL_{\textrm{\tiny JT}}[\phi] + \int_{\partial \mathscr M}\!\!\!\!\extd x L_b[\phi]
=\frac{1}{16\pi G}\int_{\mathscr M}\!\! \extd^2x  \sqrt{-g}\, X\big(R - 2 \Lambda\big)+ \frac{1}{8\pi G}\int_{\partial \mathscr M}\!\!\!\! \extd x \sqrt{-h}\, XK \\ - \frac1{8\pi G}\int_{\partial \mathscr M}\!\!\!\! \extd x \sqrt{-h}\, \Big(\frac{X}{\ell}+\frac{\ell}{2X}h^{\mu\nu}(\partial_\mu X)(\partial_\nu X)\Big)
\label{renormalized action 2}
\end{multline} 
contains the boundary metric $h_{\mu\nu}=g_{\mu\nu}-n_\mu n_\nu$, the trace of the extrinsic curvature $K=\nabla_\mu n^\mu$, and the outward pointing unit normal $n^\mu$. In the coordinates \eqref{FGgauge}-\eqref{dirichlet BC} these boundary quantities are given by $h_{\mu\nu}=-\delta_\mu^t\delta_\nu^t\,(\frac{1}{\rho^2\ell^2}+{\cal O}(1))$ (so that $\sqrt{-h}=\frac{1}{\rho\ell}+{\cal O}(\rho)$), $K=\frac1\ell+{\cal O}(\rho^2)$, and $n^\mu=-\delta^\mu_\rho\,\frac\rho\ell$. The action \eqref{renormalized action 2} admits a well-defined variational principle, provided one freezes the zero mode of $1/X^+$. Imposing this condition does not lead to any constraint on the symmetry parameters, see \eqref{variation zero mode}. 

The variation of the renormalized Lagrangian in \eqref{renormalized action 2},
\begin{align}
    \delta (L_{\textrm{\tiny JT}}[\phi] + \extd L_b[\phi]) &= \frac{\delta L_{\textrm{\tiny JT}}[\phi]}{\delta \phi^i} \delta \phi^i + \partial_\mu \Theta^\mu_{\textrm{\tiny ren}}[\phi; \delta \phi] \\ \Theta^\mu_{\textrm{\tiny ren}}[\phi; \delta \phi] &= \Theta^\mu_{\textrm{\tiny JT}}  [\phi; \delta \phi ] + \partial_\nu Y^{\mu\nu}  [ \phi; \delta \phi]  - \delta L_b[\phi]  
\end{align}
yields the renormalized presymplectic potential $\Theta_{\textrm{\tiny ren}}$, where we followed the Comp\`ere--Marolf prescription \cite{Compere:2008us} of identifying $Y$ as the symplectic potential associated to $L_b$.  The contribution to the presymplectic potential coming from the bulk JT Lagrangian 
\begin{equation}
\Theta_{\textrm{\tiny JT}}^\mu [ \phi; \delta \phi ] =   X {{\Theta}}_{\textrm{\tiny EH}}^\mu [g; \delta g] + \frac{\sqrt{-g}}{16\pi G} [- (\delta g)^{\mu \nu} \nabla_\nu X + (\delta g)^\nu_\nu \nabla^\mu X     ]   \label{presymplectic}
\end{equation}  
contains ${\Theta}_{\textrm{\tiny EH}}^\mu [g; \delta g]= \frac{\sqrt{-g}}{16\pi G} [ \nabla_\nu (\delta g)^{\mu\nu} - \nabla^\mu (\delta g)^\nu_\nu]$ and expands as
\begin{align}
   \Theta_{\textrm{\tiny JT}}^\rho [g; \delta g] & =\frac{\delta \goo X_+}{8 \pi  G }+\mathcal O\li \rho\ri \qquad\qquad  \Theta_{\textrm{\tiny JT}}^t [g; \delta g] =\mathcal O\li \rho^5\ri\,.
\end{align} 
The corner contribution 
\begin{equation}
    Y^{\rho t} [\phi; \delta \phi]= -\frac{\ell}{8\pi G X} \sqrt{-\gamma}h^{tt} \delta X \partial_tX=  \frac{\ell^2 \delta X^+ (X^+)'}{8 \pi G X^+}+\mathcal O(\rho^2)
    \label{corner contribution}
\end{equation}
comes from the boundary kinetic term for the dilaton.

We now add the matter sector given by a free scalar field
\begin{equation}
  S_{m}=\int_{\mathscr M}\!\!\extd^2 x  L_{m}[\psi, g]=\frac{1}{2}\int_{\mathscr M}\!\!\extd^2 x \sqrt{-g}\,g^{\mu\nu}(\partial_\mu\psi)(\partial_\nu\psi)
    \label{matter lagrangian scalar}
\end{equation}
where we chose a vanishing boundary action. The variation of the matter Lagrangian
\begin{equation}
\delta L_{m}[\psi, g]= \frac{\delta L_{m}[\psi, g]}{\delta \psi} \delta \Psi +\frac{\delta L_{m}[\psi, g]}{\delta g_{\mu\nu}} \delta g_{\mu \nu}  + \partial_\mu \Theta^\mu_m [\psi, g; \delta \psi]
   \label{variation matter L}
\end{equation} 
yields a contribution to the presymplectic potential,
\begin{equation}
    \Theta_m^\mu [\psi, g; \delta \psi] = \sqrt{-g}\, \delta \psi g^{\mu\nu}\partial_\nu \psi 
\end{equation}
that expands as
\begin{equation}
    \Theta_m^\rho [\psi, g; \delta \psi] = \frac1{\ell^2}\,N\,\delta J +\mathcal O(\rho)\qquad\qquad  \Theta_m^t [\psi, g; \delta \psi]=-\ell^2\,J^\prime\,\delta J + \mathcal O(\rho)\,.
\end{equation}

The projection on the boundary of the total presymplectic potential, given by the sum of the gravitational sector and the matter sector, is $\sqrt{-h}/\sqrt{-g} n_\mu\Theta^\mu=-\Theta^\rho$ and reads 
\begin{equation}
-\Theta^\rho[\phi;\delta\phi]=-(\Theta_{\textrm{\tiny ren}}^\rho[\phi; \delta \phi] +  \Theta_m^\rho [\psi, g; \delta \psi])=-\frac1{\ell^2}\,N\,\delta J -\frac1{8\pi G}\,M\,\delta\left( \frac1{X^+}\right)\,.
\end{equation}
The total presymplectic current
\begin{equation}\label{eq:wrenJT}
     \omega^\mu [\phi;\delta_1 \phi, \delta_2 \phi] = \omega^\mu_{\textrm{\tiny JT}} [\phi;\delta_1 \phi, \delta_2 \phi] + \partial_\nu (\delta_2 Y^{\mu\nu}[\phi;\delta_1 \phi] - \delta_1 Y^{\mu\nu}[\phi;\delta_2 \phi])+\omega^\mu_m [\phi;\delta_1 \phi, \delta_2 \phi]
\end{equation} 
contains $\omega^\mu_{\textrm{\tiny JT}} [\phi;\delta_1 \phi, \delta_2 \phi]=\delta _2 \Theta^\mu_{\textrm{\tiny JT}}[\phi;\delta_1 \phi]-\delta _1 \Theta_{\textrm{\tiny JT}}^\mu [\phi;\delta_2 \phi]$, the corner contribution $Y$ given in \eqref{corner contribution}
and the matter contribution $\omega^\mu_m [\phi;\delta_1 \phi, \delta_2 \phi]=\delta _2 \Theta_m^\mu[\phi;\delta_1 \phi]-\delta _1 \Theta_m^\mu [\phi;\delta_2 \phi]$.

The Iyer--Wald co-dimension 2-form (a 0-form in 2d) 
\begin{equation}
  \partial_\nu k^{\mu\nu}_{\xi}[\phi; \delta \phi] = \omega_{\textrm{\tiny ren}}^\mu [\phi;\delta_\xi \phi, \delta \phi] + \omega_{m}^\mu [\phi;\delta_\xi \phi, \delta \phi]   
  \label{boundary term symplectic structure}
\end{equation}
is obtained from the presymplectic current, yielding
\begin{equation}\label{ktrdiffeo}
    k^{\rho t}_\xi [\phi; \delta \phi]\Big|_{\rho = 0} = \frac f{8\pi G}  \,\frac{\delta M}{X^+} -\frac{f}{\ell^2}\,N\,\delta J\,.
\end{equation} 
This co-dimension 2-form is non-integrable both in the gravitational and the matter sectors. Performing the change of slicing
\begin{equation}
    f=\tilde f \, X^+ \qquad \qquad \delta \tilde f=0
\end{equation}
which amounts to a field dependent redefiniton of the symmetry generators, one obtains
\begin{equation}
    k^{\rho t}_\xi [\phi; \delta \phi]\Big|_{\rho = 0}   =  \delta \left( \tilde f \, \frac1{8\pi G}M \right)-\frac{\tilde f}{\ell^2}X^+\,N\,\delta J\,.
    \label{appcodimension 2}
\end{equation} 

\section{CFT aspects of leaky JT gravity}\label{sec:holography}

In this appendix, we reinterpret some of our gravity results from a chiral CFT$_2$ perspective, where time $t$ plays the role of the holomorphic coordinate $z$ and $\rho$ the role of the holographic coordinate associated with renormalization group flow. A concise discussion contrasting the CFT$_1$ and chiral CFT$_2$ perspectives can be found in \cite{Hartman:2008dq}. To reduce clutter, we set the AdS radius to unity, $\ell=1$.

Redefining $X^+=e^\Phi$, the falloff conditions \eqref{falloff conditions} for dilaton and metric reduce to\footnote{%
The JT black hole \eqref{eq:JTsol} corresponds to constant $M$ and vanishing $\Phi$. Thus, coupling JT gravity to matter is essential for the interpretation here.}
\begin{subequations}\label{eq:appsolutiononshell}
  \begin{align}
X &= \frac1{\rho}\,e^\Phi  + \rho\,\Big(\tfrac12\,e^{-\Phi}M-\tfrac14\,e^\Phi(\Phi^\prime)^2\Big) +\mathcal O(\rho^2)\\
g_{tt} &= - \left[ \frac{1}{\rho} -\rho\,\Big(\tfrac12\,e^{-2\Phi}M + {\cal L}_{\textrm{\tiny tS}}\Big) \right]^2 \label{eq:gtt} \\
{\cal L}_{\textrm{\tiny tS}} &\equiv \frac14\,\big(\Phi^\prime\big)^2+\frac12\,\Phi'' = \frac14\,{\cal J}^2+\frac12\,{\cal J}^\prime
\end{align}
\end{subequations}
It is suggestive to interpret the expression ${\cal L}_{\textrm{\tiny tS}}$ as a (twisted) Sugawara stress-energy tensor associated with a current ${\cal J}=\Phi^\prime$ of a scalar field $\Phi$. If this interpretation was correct, then expressions like ${\cal V}_{(h)}=e^{-h\Phi}$ should be vertex operators of conformal weight $h$. 

To verify whether this interpretation makes sense, we reconsider the transformation laws \eqref{variations solution space} under reparametrization symmetry at the boundary. The results
\begin{subequations}
\label{eq:trafos}
\begin{align}
    \delta_f \Phi &=f\,\Phi^\prime-f'  \\
    \delta_f {\cal J} &= f\,{\cal J}^\prime + f^\prime\,{\cal J} - f''\\
     \delta_f {\cal L}_{\textrm{\tiny tS}} &= f\,{\cal L}_{\textrm{\tiny tS}}^\prime + 2f^\prime\,{\cal L}_{\textrm{\tiny tS}} - \frac12\,f'''\\
     \delta_f {\cal V}_{(h)} &= f\,{\cal V}^\prime + h\,f^\prime\,{\cal V}
\end{align} 
are compatible with the suggestive CFT interpretation above: $\Phi$ is an anomalous scalar field (transforming like a Liouville field or like entanglement entropy, see, e.g., \cite{Wall:2011kb, Grumiller:2019tyl, Fiorucci:2023lpb}), ${\cal J}$ is an anomalous current (transforming with a twist cocycle, see, e.g., \cite{Afshar:2015wjm, Afshar:2019axx, Afshar:2021qvi}), ${\cal L}_{\textrm{\tiny tS}}$ is an anomalous stress-energy tensor (transforming with the Schwarzian derivative, see, e.g., \cite{Brown:1986nw, Alekseev:1988ce, Maldacena:2016upp}), and ${\cal V}_{(h)}$ transforms like a conformal primary of weight $h$, i.e., like a vertex operator (see, e.g., \cite{diFrancesco}). In the main text, we indirectly used the last fact in \eqref{variation zero mode} to conclude that the zero mode of $1/X^+={\cal V}_{(1)}$ does not transform under reparametrizations.

Finally, to make this appendix self-contained, we list again the transformation properties of some of the remaining quantities, in particular, the mass function $M$ and the modes $J$ and $N$ of the scalar field:
\begin{align}
    \delta_f M &= f\,M^\prime\\
    \delta_f J &= f\,J^\prime\\
    \delta_f N &= f\,N^\prime + f^\prime\,N\\
    \delta_f\big(e^{-2\Phi}M\big) &= f\,\big(e^{-2\Phi}M\big)^\prime+2f^\prime\,\big(e^{-2\Phi}M\big)\\
    \delta_f T_{\mu\nu}\big|_{\rho=0} &= f\,T_{\mu\nu}^\prime\big|_{\rho=0} + 2f^\prime\,T_{\mu\nu}\big|_{\rho=0} \label{eq:Ttrafo}\\
    \delta_f\big(NJ^\prime e^\Phi\big) &= f\,\big(NJ^\prime e^\Phi\big)^\prime + f^\prime\,\big(NJ^\prime e^\Phi\big)\\
     \delta_f M^\prime &= f\,M''+f^\prime\,M^\prime
\end{align}
\end{subequations} 
Under reparametrization symmetry at the boundary, $J$ and $M$ transform as scalars, $N$ as a weight-1 conformal primary, and $e^{-2\Phi}M$ as a weight-2 conformal primary (without anomalous term). Plugging these results into the leading order expressions of the bulk stress-energy tensor \eqref{eq:bulkT} reveals that each of its leading order components from a boundary perspective is a weight-2 primary (without anomalous term), see \eqref{eq:Ttrafo}. The last two expressions show that the flux-balance law \eqref{mass change} is consistent with the conformal transformation properties, i.e., the term $NJ^\prime e^\Phi$ transforms precisely as $M^\prime$ does. 

Thus, from a chiral CFT$_2$ perspective, the subleading term in the metric component $g_{tt}$ \eqref{eq:gtt} is the sum of a weight-2 primary, $e^{-2\Phi}M$, and a weight-2 quasiprimary, the twisted Sugawara stress-energy tensor ${\cal L}_{\textrm{\tiny tS}}$. The latter requires non-constant $\Phi$ and hence the presence of matter.

\section{Dilaton-dependent measure}\label{sec:6}

In this appendix, we consider yet-another alternative to the main text, where we deform the path integral measure using the dilaton field to remove the divergence in the definition of the diffeomorphism invariant path integral measure \eqref{diffeo invariance}. 
\begin{equation}
    1 = \int D\psi \, \exp\Big[ - \int \extd^2 x\sqrt{-g}\, X^\eta \psi^2 \Big]
    \label{measure with X}
\end{equation} 
While this measure is not natural for minimally coupled matter fields (unless $\eta=0$), it is diffeomorphism invariant for any $\eta$. In cases where the scalar field in 2d emerges from dimensional reduction of some higher-dimensional theory, we typically have $\eta=1$ (though in this case, the same linear factor of the dilaton would also appear in front of the kinetic term for the scalar field in the action).

Considering our falloff conditions \eqref{falloff conditions}, the divergence problems in the definition of the measure discussed in \eqref{divergences in the measure} are solved by taking 
\begin{equation}
    \eta \le -2\,.
\end{equation} 
We define 
\begin{equation}
    \tilde{\psi} = X^{\frac{\eta}{2}} \psi \qquad \qquad D \tilde \psi = X^{\frac{\eta}{2}} D \psi 
    \label{change of variables}
\end{equation} 
so that \eqref{measure with X} reads as
\begin{equation}
    1 = X^{-\frac{\eta}{2}} \int D\tilde\psi \, \exp\Big[ - \int \extd^2 x \sqrt{-g}\, \tilde \psi^2 \Big]\,.
    \label{measure with X 2}
\end{equation} 
Performing the change of variables \eqref{change of variables} in \eqref{path integral}, the path integral 
\begin{equation}
    Z_m = X^{-\frac{\eta}{2}}  \int D \tilde \psi \, \exp\Big[ - \frac{1}{2} \int \extd x \extd y \, \tilde \psi(x) \sqrt{-g} \tilde{K}(x,y)  \tilde \psi (y) \Big]
\end{equation} 
involves the kernel
\begin{equation}
    \tilde{K}(x,y) = X(x)^{-\frac{\eta}{2}} \Box   X(y)^{-\frac{\eta}{2}} \,.
    \label{tilde K}
\end{equation} 
Using the dilaton-dependent diffeomorphism invariant measure \eqref{measure with X 2}, one can integrate over $\tilde{\psi}$ to obtain
\begin{equation}
     Z_m = (\det \tilde{K})^{-\frac{1}{2}}\,.
     \label{path integral X}
\end{equation} 
To proceed, one would need to regularize this result using, e.g., heat kernel methods \cite{Vassilevich:2003xt}. 

The Gauss law is violated only if there is a diffeomorphism anomaly arising at one loop. Since the classical action and the path integral measure \eqref{measure with X} are both diffeomorphism invariant, the one-loop effective theory is diffeomorphism invariant. Therefore, there is no violation of the Gauss law, which implies that the Noether identities associated with diffeomorphisms are satisfied. It is still interesting to derive the Noether identities for the one-loop effective theory obtained after integrating out the scalar field $\psi$ using the dilaton-dependent measure \eqref{measure with X}: $L[\phi] = L_{\textrm{\tiny JT}}[g,X] + L_{\text{eff}}[g,X]$ with $\phi = (X, g)$ and $L_{\text{eff}}[g,X]$ is the effective Lagrangian obtained after using the path integral measure \eqref{measure with X} for the matter sector. Notably, the effective Lagrangian now depends on the dilaton field $X$. Repeating the steps in \eqref{step 1}, we have
\begin{equation}
    \delta_\xi \phi^i \frac{\delta L}{\delta \phi^i} = \xi^\mu \partial_\mu X \frac{\delta L}{\delta X} + 2 \nabla_\mu\xi_\nu \frac{\delta L}{\delta g_{\mu\nu}} 
    = \xi^\mu N_\mu[\phi] + \partial_\mu S^\mu_\xi[\phi]
\end{equation} 
where, as in section \ref{sec:5}, $N_\mu[\phi]$ denotes the Noether identities and $S^\mu_\xi[\phi]$ is the weakly vanishing Noether current, given by
\begin{equation}
    N_\mu [\phi] = \partial_\mu X \frac{\delta L_{\text{eff}}}{\delta X} + \sqrt{-g}\, \nabla^\nu \langle T_{\mu \nu} \rangle \qquad \qquad S^\mu_\xi [\phi] = \xi_\nu \frac{\delta L}{\delta g_{\mu\nu}} \,.
\end{equation} 
Diffeomorphism invariance implies $N_\mu = 0$ off-shell, i.e.,
\begin{equation}
    \boxed{\phantom{\big(}
    \nabla^\nu \langle T_{\mu \nu} \rangle = - (\partial_\mu X) \frac{1}{\sqrt{-g}} \frac{\delta L_{\text{eff}}}{\delta X}\neq 0\,.
    \phantom{\Big)}} 
    \label{breaking T}
\end{equation} 
This matches with (6.40) of the review \cite{Grumiller:2002nm}. Classically, we had $\nabla^\nu T_{\mu\nu} = 0$ on-shell for the matter sector. Semi-classically, the stress-energy tensor is not conserved due to the dilaton field. However, as opposed to the situation in section \ref{sec:Local and Weyl invariant measure}, this anomalous symmetry breaking does not violate (and indeed, it is even essential to satisfy) the Gauss law.

\section{More on Gauss law breaking}\label{app:IyerWaldeffectiveaction}

In this appendix, we provide an alternative derivation for the breaking of the gravitational Gauss law \eqref{app:IyerWaldeffectiveaction} using the Iyer--Wald covariant phase space methods \cite{Iyer:1994ys}. We also display more explicit formulae for this breaking. 

First, we can rewrite $\Gamma_{\text{eff}}^W[\gamma,\chi]$ in Equation \eqref{Weyleffectiveaction} to make the metric $g$ appear explicitly:
\begin{equation}
\Gamma_{\text{eff}}^W[g,\chi] = - \frac{c}{24 \pi} \int \extd^2 x \,\sqrt{g}  \left[g^{\mu\nu}(\partial_\mu \chi) (\partial_\nu \chi) + \chi \left( R+\Box_g\ln\sqrt g\right)
\right]\,.
\end{equation} 
Similarly, the stress-energy tensor \eqref{stress tensor Weyl} in terms of the metric reads as
\begin{align} 
   T^W_{\mu\nu} = \frac{2}{\sqrt{g}}\frac{\delta \Gamma^{W}_{\text{eff}}[g,\chi]}{\delta g^{\mu\nu}} =& -\frac{c}{12\pi}\left((\partial_\mu \chi)(\partial_\nu\chi)-\frac12 g_{\mu\nu} (\partial \chi)^2-\nabla_\mu \nabla_\nu\chi + {\frac12}\Box_g\chi g_{\mu\nu} \right) \\
& -\frac{c}{12\pi}\left( -\frac12 (\partial_\mu\chi)(\partial_\nu \ln\sqrt{g}) -\frac12 (\partial_\nu\chi)(\partial_\mu \ln\sqrt{g}) + \frac12 g_{\mu\nu}g^{\sigma\lambda}(\partial_\sigma\chi)(\partial_\lambda \ln\sqrt{g})\right)\nonumber
\end{align} 
Under diffeomorphisms, the transformation laws
\begin{subequations}
    \label{variations action}
\begin{align}
    \delta_\xi \gamma_{\mu\nu}& =\xi^\sigma\partial_\sigma\li \gamma_{\mu\nu}  \ri + \gamma_{\mu\sigma}\partial_\nu\xi^\sigma+ \gamma_{\nu\sigma}\partial_\mu\xi^\sigma -\li\partial\cdot \xi\ri\gamma_{\mu\nu}\\ 
    \delta_\xi \mathcal R& = \partial_\mu\li \xi^\mu \mathcal R\ri+\Box_\gamma (\partial\cdot\xi) \\
    \delta_\xi T_{\mu\nu}^W&=\xi^\sigma\partial_\sigma T_{\mu\nu}^W +T^W_{\mu\sigma}\partial_\nu\xi^\sigma+ T^W_{\nu\sigma}\partial_\mu\xi^\sigma \nonumber\\
    & +\frac{c}{24 \pi}\li \partial_\mu \li\partial\cdot\xi\ri \partial_\nu \chi+\partial_\nu \li\partial\cdot\xi\ri\partial_\mu \chi -\gamma_{\mu\nu} \gamma^{\sigma\lambda}\partial_\sigma \li\partial\cdot\xi\ri\partial_\lambda \chi \ri 
\end{align}
\end{subequations} 
have additional terms with respect to the standard Lie derivative acting on tensors, which all involve the divergence of $\xi$, $\partial\cdot\xi \equiv \partial_\mu \xi^\mu$ and are only there for non-area preserving diffeomorphisms. In particular, the inhomogeneous terms in the transformation of $T^W_{\mu \nu}$ are reminiscent of the Schwarzian derivative in a 2d CFT or similar anomalies in the sense that they depend only on derivatives of the transformation parameter $\xi$ and on background quantities ($\gamma$ and $\chi$), but not on the transformed field. Moreover, these anomalous terms scale linearly with the central charge $c$. 

The Euler--Lagrange derivatives of the one-loop effective Lagrangian
$L=L_{\textrm{\tiny JT}}+L_{\text{eff}}^W$, 
\begin{align}
\frac{\delta L}{\delta g_{\mu\nu}}&= \frac{\sqrt{-g}}{16\pi G} \left(\nabla^\mu \nabla^\nu X - g^{\mu\nu} \nabla^2 X - g^{\mu\nu} \Lambda X\right)  -\frac{\sqrt{-g}}2  T_W^{\mu\nu} \\
 \frac{\delta L}{\delta X}&=\frac{\sqrt{-g}}{16\pi G} \left( R - 2 \Lambda\right) , \\  
 \frac{\delta L}{\delta \chi}&=\frac{c}{12\pi}\li\Box_\gamma \chi -\frac12 \mathcal R\ri
\label{EOM eff W}
\end{align} 
together with the variations \eqref{variations action}, yield
\begin{subequations}\label{eq:transfoofELWeyl}
\begin{align}
\delta_\xi\li \frac{\delta L}{\delta X}\delta X \ri 
&=\partial_\sigma\li \xi^\sigma \frac{\delta L}{\delta X}\delta X \ri\\
\delta_\xi\li \frac{\delta L}{\delta \chi}\delta \chi \ri &=\partial_\sigma\li \xi^\sigma \frac{\delta L}{\delta \chi}\delta \chi \ri - \frac{c}{24\pi} \sqrt{g}\Box_g\li\partial\cdot\xi\ri \delta\chi\\
\delta_\xi\li \frac{\delta L}{\delta g_{\mu\nu}}\delta g_{\mu\nu} \ri 
&=\partial_\sigma\li \xi^\sigma \frac{\delta L}{\delta g_{\mu\nu}}\delta g_{\mu\nu} \ri +\frac{c}{24\pi} \partial_\mu(\partial\cdot\xi)\partial_\nu \chi \delta (\sqrt g g^{\mu\nu})\,.
\end{align}
\end{subequations} 
Again, the terms violating the diffeomorphism covariance all involve $\partial\cdot\xi$. 

With these formulae, we are now ready to derive the breaking of the Gauss law in the Iyer--Wald formalism. We follow the steps of \cite{McNees:2023tus} by keeping track of the anomalous terms. We start by computing 
\begin{equation}
    \delta L= \frac{\delta L}{\delta \phi}\delta\phi+\partial_\mu\Theta^\mu[\phi;\delta\phi]
\end{equation} 
so that
\eq{
\delta \delta_\xi L = \delta\left(\frac{\delta L}{\delta \phi}\delta_\xi\phi\right)+\partial_\mu\delta \Theta^\mu[\phi;\delta_\xi \phi]\qquad\qquad \delta_\xi \delta L = \delta_\xi\left(\frac{\delta L}{\delta \phi}\delta\phi\right)+\partial_\mu\delta_\xi \Theta^\mu[\phi;\delta \phi]\,.
}{eq:align}
Then we use $\omega^\mu[\phi;\delta_\xi\phi,\delta \phi]=\delta \Theta^\mu[\phi;\delta_\xi \phi]-\delta_\xi \Theta^\mu[\phi;\delta \phi]$ and \eqref{step 1} to write
\begin{equation}
    [\delta,\delta_\xi]L=\partial_\mu\omega^\mu[\phi;\delta_\xi\phi,\delta \phi] +\delta\left(\partial_\mu S^\mu_\xi[\phi]+\xi^\mu N_\mu[\phi]\right)-\delta_\xi\left(\frac{\delta L}{\delta \phi }\delta\phi\right)=0\,.
\end{equation}
When the Euler Lagrange derivatives do not transform as densities, $\delta_\xi(\delta L/\delta \phi \delta \phi)=\partial_\mu(\xi^\mu \delta L/\delta \phi \delta \phi)+A_\xi[\phi;\delta\phi]$, we have
\begin{align}
\partial_\mu\left(\omega^\mu[\phi;\delta_\xi\phi,\delta \phi] +\delta S_\xi^\mu[\phi] -\xi^\mu \frac{\delta L}{\delta \phi}\delta\phi\right) +\xi^\mu \delta N_\mu[\phi]-A_\xi [\phi;\delta\phi]=0\,.
\label{breaking IW Gauss law}
\end{align}
For our case, $A_\xi$ can be read off from \eqref{eq:transfoofELWeyl},
\begin{align}
A_\xi[\phi;\delta\phi]&=\frac{c}{24\pi}  \left(
- \sqrt{g}\Box_g\li\partial\cdot\xi\ri \delta\chi
+\partial_\mu(\partial\cdot\xi)\partial_\nu \chi \delta (\sqrt g g^{\mu\nu}) \right)\nonumber \\
&=-\frac{c}{24\pi} \left((\partial\cdot\xi) \delta( \sqrt{g}\Box_g \chi ) +\partial_\mu\left( \sqrt{g} g^{\mu\nu}\partial_\nu(\partial\cdot \xi)  \delta\chi-(\partial\cdot \xi) \delta \li \sqrt{g} g^{\mu\nu}\partial_\nu \chi \ri  \right) \right)\label{anomaly}
\end{align}
where we used 
\begin{align}
 \sqrt{g}\Box_g \delta \chi&=\delta (\sqrt{g}\Box_g \chi)-g^{\mu\nu}\partial_\nu \chi \partial_\mu\delta \sqrt{g}-\Box_g\chi\delta \sqrt{g}-\partial_\mu\li\sqrt{g} \partial_\nu\chi \delta g^{\mu\nu}\ri\,.
\end{align}
Then the two last terms of \eqref{breaking IW Gauss law} can be rewritten as a total derivative,
\begin{align}
 \xi^\mu \delta N_\mu[\phi]-  A_\xi[\phi;\delta\phi] & =\frac{c}{24\pi}\partial_\mu\left( \xi^\mu \delta( \sqrt{g}\Box_g\chi ) + \sqrt{g} g^{\mu\nu}\partial_\nu(\partial\cdot \xi)  \delta\chi-(\partial\cdot \xi) \delta \li \sqrt{-g} g^{\mu\nu}\partial_\nu \chi \ri  \right) \,.
\end{align}
On-shell, \eqref{breaking IW Gauss law} reads as
\begin{align}
&  \partial_\mu \omega^\mu[\phi;\delta_\xi\phi,\delta \phi] =  - \frac{c}{24\pi}\partial_\mu \left(\sqrt{g} g^{\mu\nu}\partial_\nu(\partial\cdot \xi)  \delta\chi  
 + \delta \left[\sqrt{g} \li \xi^\mu \Box_g\chi -(\partial\cdot \xi)   g^{\mu\nu}\partial_\nu \chi  \ri \right]\right) \,.\label{derivative of Gauss law}
\end{align} 
If there was no anomaly, one would expect to find a formula of the form \eqref{boundary term symplectic structure}, i.e., $\omega^\mu[\phi;\delta_\xi\phi,\delta \phi] = \partial_\nu k^{\mu \nu}_\xi$, stating that the gravitational charges are co-dimension 2. In particular, the derivative of this equation would lead to $\partial_\mu \omega^\mu[\phi;\delta_\xi\phi,\delta \phi] = 0$. However, the right-hand side \eqref{derivative of Gauss law} unveils the presence of an anomalous term breaking the gravitational Gauss law, confirming our result \eqref{breaking Gauss law} using this alternative method.  

\newcommand{\tone}{I_1} 
\newcommand{\ttwo}{I_2} 
\newcommand{\tthree}{I_3} 

Now, as a final check, we verify that the breaking terms found in \eqref{breaking Gauss law} are compatible with those obtained in \eqref{derivative of Gauss law}. To do so, we take the derivative of \eqref{breaking Gauss law} and compare it with the above formula. The breaking term in \eqref{breaking Gauss law} reads as   
\begin{align}
    I^n_{\delta \phi} \left[\xi^\mu   \li\frac{c}{24\pi}\partial_\mu \li\sqrt{-g}\, \Box_g\chi \ri  \ri (\extd^2x)\right] 
\end{align}
We write 
\begin{equation}
  (  \xi^\mu\partial_\mu \li\sqrt{-g}\, \Box_g\chi \ri)  (\extd^2x)= \big(\tone+\ttwo+\tthree\big) (\extd^2x)
    \label{decomposition}
\end{equation}
with 
\begin{align}
    \tone& = \partial_\sigma(\partial \cdot \xi)\sqrt{g}\,g^{\sigma\nu}\partial_\nu \chi \\
    \ttwo&= -\partial_\mu\left(  \partial_\sigma \xi^\mu  \sqrt{g}g^{\sigma\nu} \partial_\nu \chi + (\partial \cdot \xi) \sqrt{g}g^{\mu\nu} \partial_\nu \chi \right)=:\partial_\mu B^\mu \\
    \tthree&=\partial_\mu\partial_\nu\li \xi^\mu\sqrt{g}g^{\sigma\nu}  \partial_\nu X \ri\,.
\end{align} 
Following the conventions in Equation (B.16) of \cite{Ruzziconi:2019pzd} for the homotopy operator, we have
\begin{align}
 ( I^n_{\delta \phi} \tone (\extd^2x))^\mu& =\partial_\sigma(\partial \cdot \xi)\sqrt{g}g^{\sigma\mu}\delta\chi  \\ \nonumber
 ( I^n_{\delta \phi} \ttwo (\extd^2x))^\mu& 
  = \delta g_{ab} \frac{\partial}{\partial g_{ab} } B^\mu + \delta \chi  \frac{\partial}{\partial \chi } B^\mu -\delta \chi \partial_\sigma  \frac{\partial}{\partial  (\partial_\sigma\chi ) }  B+\partial_\sigma\left(\delta^{(\sigma}_\nu \frac{\delta B^\nu}{\partial_{\mu)}\chi}\delta\chi \right) \\
&= - \delta \left(\sqrt{g} \partial_\sigma\xi^\mu g^{\sigma\lambda}\partial_\lambda\chi +(\partial\cdot \xi)   g^{\mu\nu}\partial_\nu \chi   \right) +\partial_\sigma\left(\sqrt{g}\partial_\lambda\xi^{[\mu} g^{\sigma]\lambda} \delta\chi\right)
 \\
( I^n_{\delta \phi} \tthree (\extd^2x))^\mu& = \delta \partial_\sigma \left( \sqrt{g} \xi^{(\sigma}g^{\mu)\lambda}\partial_\lambda \chi \right) +\frac13\partial_{\sigma\lambda}\left(\sqrt{g}  \xi^{[\lambda} g^{\mu]\sigma } \delta \chi \right)
\end{align} 
Then we have
\begin{align}\nonumber
    I^n_{\delta \phi} &\left[\xi^\mu   \li\frac{c}{24\pi}\partial_\mu \li\sqrt{-g}\, \Box_g\chi \ri  \ri (\extd^2x)\right]  \\\nonumber
    &=\big( \partial_\sigma(\partial \cdot \xi)\sqrt{g}g^{\sigma\mu}\delta\chi + \delta \left(\sqrt{g} \li \xi^\mu \Box_g\chi -(\partial\cdot \xi)   g^{\mu\nu}\partial_\nu \chi  \ri \right)\\ \label{expression breaking gauss law}
&+\partial_\sigma\left(\sqrt{g}\partial_\lambda\xi^{[\mu} g^{\sigma]\lambda} \delta\chi + \delta\left(\sqrt{g} \xi^{[\sigma}g^{\mu]\lambda}\partial_\lambda \chi\right)\right)
+\frac13\partial_{\sigma\lambda}\left(\sqrt{g}  \xi^{[\lambda} g^{\mu]\sigma } \delta \chi \right) \big)(\extd^1x)_\mu 
\end{align}
Finally, taking the derivative of \eqref{breaking Gauss law} leads to
\begin{align}
    \partial_\mu W^\mu[\phi; \delta_\xi \phi, \delta \phi] =  + \frac{c}{24\pi}\partial_\mu \left(\sqrt{g} g^{\mu\nu}\partial_\nu(\partial\cdot \xi)  \delta\chi  
 + \delta \left(\sqrt{g} \li \xi^\mu \Box_g\chi -(\partial\cdot \xi)   g^{\mu\nu}\partial_\nu \chi  \ri \right)\right)
\end{align}
which matches exactly with \eqref{breaking IW Gauss law}, using the following relation: $\partial_\mu\omega^\mu[\phi; \delta_\xi \phi, \delta \phi]  = -\partial_\mu W^\mu[\phi; \delta_\xi \phi, \delta \phi]$ (see, e.g., \cite{Compere:2007az,Ruzziconi:2019pzd}).


\bibliographystyle{fullsort.bst}
\bibliography{references}


\end{document}